\documentclass[12pt]{iopart}
\usepackage{iopams}  
\usepackage{graphicx}  
\usepackage{color}  
\usepackage{epsfig}
\usepackage{bm}
\usepackage{url}

\usepackage[mathscr]{eucal}
\usepackage{latexsym}
\usepackage{stmaryrd}
\RequirePackage{geometry}
\RequirePackage{fancybox}
\RequirePackage{pstricks,soul,colortbl,color}
\RequirePackage{fancyhdr}
\RequirePackage{graphicx}
\usepackage{texdraw}
\usepackage{mathrsfs}
\RequirePackage{pstricks,soul,colortbl,color}
\usepackage{times}
\usepackage{cite}
\RequirePackage{amsfonts,amsthm,dsfont,latexsym,amssymb}
\usepackage{amssymb, amsthm, amsbsy, amsfonts, amsthm, paralist}

\newcommand{\bbC}{\mathds{C}}

\begin{document}

\title{Study of multi black hole and ring singularity\\
apparent horizons}

\author{Gabriela Jaramillo and Carlos O. Lousto}

\address{Center for Computational Relativity and Gravitation, and
School of Mathematical Sciences, 
Rochester Institute of Technology, Rochester, New York 14623, USA}
\ead{jara0025@umn.edu,lousto@astro.rit.edu,}

\begin{abstract}
We study critical black hole separations for the formation of a 
common apparent horizon in systems of $N$ - black holes in a time
symmetric configuration. We study in detail the aligned
equal mass cases for $N=2,3,4,5$,
and relate them to the unequal mass binary black hole case.
We then study the apparent horizon of the time symmetric initial 
geometry of a ring singularity of different radii. The apparent horizon is
used as indicative of the location of the event horizon in an effort to predict
a critical ring radius that would generate an event horizon of toroidal
topology. We found that a good estimate for this ring critical radius is 
$20/(3\pi) M$. We briefly discuss the connection of this two cases through
a discrete black hole 'necklace' configuration.
\end{abstract}

\submitto{\CQG}
\maketitle

\section{Introduction}
%

The field of Numerical Relativity (NR) has progressed rapidly
since the 2005 breakthroughs \cite{Pretorius:2005gq,Campanelli:2005dd,Baker:2005vv}. 
Naturally, the first application
of these techniques was to solve the
non-linear dynamics of the inspiral, merger, and
ringdown of an orbiting black-hole binary (BHB). The
computation of the gravitational waveforms generated
by such systems is of utter interest
for gravitational wave observatories such as LIGO, VIRGO and LISA.
The computation of the merger of BHB is also of astrophysical interest.
In particular, the discovery of very large recoil velocities 
\cite{Campanelli:2007ew,Campanelli:2007cga} 
acquired by the final remnant of the merger has attracted
lots of interest among astrophysicists.

There are other very important applications of the new NR techniques.
Those lie in the field of Mathematical Relativity. Some few examples
are given by the studies of the geometry of maximally
spinning black holes \cite{Dain:2008ck,Lovelace:2008tw} 
behaving like $1/\sqrt{r}$ rather than $1/r$ for submaximal
near the puncture. The late time behavior of the metric conformal factor
in the 'moving puncture' approach also behavs like
\cite{Hannam:2006vv,Brown:2009ki,Immerman:2009ns,Hannam:2009ib} $1/\sqrt{r}$. 
Numerical simulations started to test
the 'no hair' theorem \cite{Campanelli:2008dv} 
and the 'cosmic censorship' conjecture \cite{Campanelli:2006uy,Campanelli:2006fy}. 
The isolated horizon formalism \cite{Dreyer:2002mx} has been implemented
numerically and validated in highly nonlinear regimes. In particular
a new proposal to measure quasilocally linear momenta from the horizon
deformation of black holes has been put forward in \cite{Krishnan:2007pu}.
In this paper we turn into the study of the merging of
apparent horizon of $N$-black hole systems and a ring singularity
in a time symmetric initial geometry. These studies can be used as
a guide to search for event horizons in more dynamical situations.

%

In the next subsections we review basic definitions that will help us
define and study apparent horizons for systems of $N$ black holes and
black hole rings. We start with the definition of an event horizon and
continue with apparent horizons, the equations used to find them and a
basic summary of the algorithms used in this project to solve these
equations. A follow up paper \cite{PLZ} will deal with the event horizon
studies.

In Sec. \ref{NBHs} we study systems of $N$ aligned Schwarzschild black holes in a
time-symmetric spacelike hypersurface. The equations involved and the
numerical methods used are presented and explained. Additionally, the
relationship between a system of two black holes of different mass and
systems of $N$ black holes with equal mass is explored.

In section \ref{ringBH} we take advantage of the equations used for systems
of $N$ black holes and adapt them to find the apparent horizon of a
black hole with a rings singularity of different mass (or equivalently 
keeping constant the total mass and changing the radius.) 
This allows us to confirm that they comply with the
results obtained by Galloway \cite{Galloway:2006ws} regarding the spherical
topology of apparent horizons in stationary black holes
spacetimes. Additionally, the apparent horizon is used as an
approximation to the event horizon and extrapolation is used to
determine the size of the black hole ring that would give rise to an
event horizon of toroidal topology.
We end with a discussion of the possibility of building up a
toroidal black hole with a discrete set of black holes in a
ring-like distribution.

\subsection{Definitions:}

In an asymptotically flat spacetime the \textbf{\emph{black hole
    region}} is a region from which no null curve can reach future
null infinity ($\mathscr{I^+}$ ), the boundary of this region is the
\textbf{\emph{event horizon}}. Since the black hole region only ceases
to increase when no more matter falls into it, its boundary cannot be
determined until all interactions between the black hole and the
surrounding matter are over. This means that in order to find the
event horizon one must complete a full simulation of the evolution of
the black hole. A more local structure such as an apparent horizon
provides a way to overcome this requirement. Since the existence of an
apparent horizon is a necessary condition for the existence of an
event horizon and because an apparent horizon will always lie inside
an event horizon, these objects have become very useful in numerical
relativity. In fact there are certain algorithms that make use of
\textquotedblleft horizon pretracking", more fully described in
\cite{Thornburg:2006zb}, or \textquotedblleft black hole excision techniques"
\cite[page 214]{AlcubierreBook2008}, where the goal is to find the apparent
horizons as soon as they appear in a simulation in order to remove the
singularity and measure the mass and angular momentum of the black
hole .

Now, there are certain cases where the problem of finding the event
horizon can be simplified. For example when we are working in
stationary, asymptotically flat spacetimes, the event horizon is a
null three surface $H$, tangent to one or more Killing vector fields
of the full spacetime. These types of horizons are formally known as
\textbf{\emph{Killing horizons}}.  On the other hand, If the Killing
vector field is not of the full spacetime, but rather of some
neighborhood of the null three surface H, then the Killing Horizon
does not coincide with the event horizon, but it is close to it
\cite{Booth:2005qc}.

Still, apparent horizons are, for the most part, the best way to
locate a black hole. But before defining what exactly is an apparent
horizon we need to define first a trapped surface. Booth describes for
Kerr-Newman black holes the \textbf{\emph{trapped surface}} as a
closed two-surface $S$ with the property that all null geodesics that
are normal to the surface and are pointing forward in time have
negative expansion everywhere \cite{Booth:2005qc}:
 
 \begin{equation}
 \theta(l) = q^{ab}\nabla_a l_b <0 \quad \textrm{and} \quad  \theta(n) = q^{ab}\nabla_a n_b <0
 \end{equation}
 
Here $q_{ab} = g_{ab} + l_an_b + l_bn_a$ is the two metric induced on
$S$ and $l^a$ , $n^a$ are the outward, inward pointing null directions
with $l\cdot n = -1$.
 
Then, given a spacetime that can be foliated into hypersurfaces
$\Sigma_t$, a point $q \in \Sigma_t$ is said to be trapped if it lies
on a trapped surface of $\Sigma_t$. An \textbf{\emph{apparent
    horizon}} is the boundary of the union of all trapped points. When
this boundary is differentiable, the apparent horizon is a \textbf{
  \emph{marginally outer trapped surface, MOT} } ( $\theta(l)=0$). In
other words, the apparent horizon is a trapped surface in which light
rays have zero expansion in the null directions that are normal to the
surface. It is this definition of an apparent horizon that has helped
develop algorithms to find it. The one used in this project is based
on the description of " shooting algorithms in axisymmetry" by
Thornburg \cite{Thornburg:2006zb} and Bishop \cite{Bishop82,Bishop84}.

In the process of finding apparent horizons for systems of $N$ black
holes, we find that there is a certain distance between black holes
that creates a common apparent horizon. We will refer to this distance
as the \textbf{\emph{critical separation $a_c$}}. For example, if two
black holes are at a distance $a_c$ or less from each other then a
common apparent horizon will form between them. On the other hand, if
the two black holes are at a distance greater than $a_c$ then two
apparent horizon will form, each surrounding one of the two black
holes.

\subsection{Motivation}

As mentioned before, the event horizon represents the true boundary of
the black hole. However, in order to find it we need to know which
outgoing null rays escape to infinity and which ones do not. They only
way to achieve this is by knowing the entire history of the
spacetime. This requires a complete simulation of the evolution of the
black hole. That is why locating the apparent horizon is so important
in Numerical Relativity. It represents a local boundary for the black
hole region and provides physical information about the black hole
such as mass and angular momentum. They are also used in numerical
simulations to locate the black holes so that black hole excision
techniques can be used. It is for these reasons that we have focused
our research in apparent horizons.

On the other hand, during the past few years system of three black
holes have been studied 
\cite{Campanelli:2005kr,Campanelli:2007ea,Lousto:2007ji,Lousto:2007rj}.
Moreover, the good probability of finding
systems of three or even more black holes \cite{Gultekin:2003xd} in globular
clusters has motivated us to consider methods for the general case of
$N$ black holes. As a starting point for more in depth future research
we have restricted ourselves to the stationary axisymmetric case.

As an extension to the methods developed in the the study of $N$ black
holes we also consider black hole rings. The paper
\cite{Shapiro95a} proving the existence of toroidal event
horizons in rotating clusters of toroidal configuration motivated us
to study these black hole rings. In this case we have considered the
apparent horizon as an approximation to the event horizon. The effects
of changing the mass of the black hole ring on the shape of the
apparent horizon are studied. The results were tabulated in order to
make a prediction about of topology of the event horizon.

\subsection{Finding Apparent Horizons}

The problem of finding an apparent horizon assuming an axisymmetric
spacetime can be reduced to solving a non linear boundary value
problem, as described in the following paragraphs. Then a numerical
method can be used to solve this boundary value problem. The following
derivation of the equations needed to find an apparent horizon in an
axisymmetric case is a summary of the methods described in \cite[pages
  221-226]{AlcubierreBook2008}, and can be found there in more detail.

Consider a spacetime manifold $M$ with metric $g_{\alpha\beta}$ and a
spacelike hypersurface $\Sigma$ in this manifold. Let $\gamma_{ij}$ be
the induced metric on the hypersurface $\Sigma$ and $K_{ij}$ be the
extrinsic curvature. Here is worth mentioning the distinction between
intrinsic curvature and extrinsic curvature. The \emph{intrinsic
  curvature} of a hypersurface comes from its internal geometry and is
given by the three dimensional Riemann tensor defined in terms of the
metric $\gamma_{ij}$. The extrinsic curvature on the other hand is
associated with the way these hypersurfaces are embedded in
spacetime. It describes how the normal vector to the surfaces changes
as its parallel transported from one point to the other. This change
is described by the extrinsic curvature tensor $K_{ij}$
\cite{AlcubierreBook2008} 69.

In this hypersurface consider a smooth 2D surface $S$ embedded in it
with a unit outward pointing normal vector $n^{\mu}$. Then the
expansion, $\theta$, of null rays which are moving in the $n^{\mu}$
direction of $S$ is given by:

\begin{equation}
\theta = \nabla_i n^i + K_{ij}n^in^j + K
\label{expansion}
\end{equation}

Where $K$ is the trace of the extrinsic curvature $K_{ij}$ and
$\nabla_i$ is the covariant derivative with respect to the metric
$\gamma_{ij}$. As mentioned before the apparent horizon is a
marginally trapped surface therefore it will be the surface for which
$\theta =0$.

If the surface is parametrized by a level set (a surface for which the
time coordinate is a constant) :

\begin{equation}
F(x^i) = 0 
\end{equation}

Then the normal vector to this surface is just the gradient of F:

\begin{equation}
n^i = \frac{\nabla^i F}{|\nabla F|}
\label{nvector}
\end{equation}

Plugging equation \ref{nvector} into equation \ref{expansion} we obtain the following:\\

\begin{equation}
\theta = \frac{\gamma^{ij}\nabla_iF \nabla_jF}{|\nabla F|} - \frac{\nabla^i F \nabla^jF\nabla_i \nabla_jF}{|\nabla F |^3} + K_{ij} \frac{\nabla^i F\nabla^j}{|\nabla F|^2 } + K
\end{equation}

Then the level set curve that satisfies $\theta=0$ would be the
apparent horizon. In the case of axisymmetric space, which is the case
we are considering, the level surface can be expressed as:

\begin{equation}
F(r,\phi) = r- h(\phi)
\end{equation}

This parameterization implies that we are considering apparent
horizons which have a center and rays leaving this center will
intersect the apparent horizon only once. In other words the
parameterization assumes that the apparent horizon has a spherical
topology. Another assumption is that the apparent horizon must be a
smooth surface. This assumption suggests that when $\phi =0$ and
$\phi=\pi$ we have $\partial_{\phi}h=0$

In his papers \cite{Bishop82}, \cite{Bishop84} Bishop assumes that the
extrinsic curvature $K_{ij}$ is zero. This simplification can be done
because we are working in a time symmetric hypersurface and so the
black holes are not moving in this time slice. Hence the equation for
the expansion reduces to:

\begin{equation}
\theta = \nabla_i n^i
\end{equation}

This implies that under these conditions the apparent horizon is an
extremal (minimal) surface. Hence, it is possible to find the apparent
horizon by finding a surface in $\Sigma$ of minimal area. This method
is described in the next section.

\section{Systems of N black holes in a line}\label{NBHs}

This section is concerned with finding the apparent horizon for
systems of $N$ black holes. First a system of two black holes of
different mass is analyzed. A table relating the mass ratio of the two
black holes and their critical separation is reproduced. Then systems
of three, four and five black holes are considered. These systems are
treated as if they contained only two black holes by grouping the
black holes adequately. The table is then used to make a prediction
about the location of the apparent horizon of these systems. These
predictions are then compared to the actual location of the apparent
horizon obtained using Bishop's equations \cite{Bishop82}. Finally a
method for finding an approximation of the apparent horizon of a
system of $N$ black holes, by representing it as a system of two black
holes of different mass, is developed.\\

\subsection{Equations}

The equations used to find the apparent horizon are presented in
references \cite{Shapiro95a}, \cite{Bishop82} and \cite{Bishop84}. A
summary of the method is given here. It was assumed that the spacelike
slice is a time-symmetric hypersurface with axial symmetry. In
cylindrical coordinates the hypersurface has the following metric:

\begin{equation} ds^2= \Psi ^4 (d\rho^2 + \rho^2 d\theta ^2 + dz^2) \end{equation}

Assuming G=c=1 and with:

\begin{equation} \Psi =  1 + \sum_i \frac{m_i}{2 R_i} \end{equation}

Here $R_i= r-r_i$ is the difference between a reference point $r =(\rho,z)$ and the location of the $i^{th}$ black hole $r_i=(\rho_i, z_i)$.\\

As mentioned in the introduction the apparent horizon is a marginally
outer trapped surface. Given the assumption that we are working in a
time symmetric hypersurface (the black holes are not moving in this
time slice) this implies that the intrinsic curvature $K_{ij}
=0$. Hence the equation for the expansion of null rays normal to the
surface is:\\

\begin{equation}
\theta = \nabla_in^i 
\end{equation}

This implies that for this particular case finding marginally trapped
surfaces is equal to finding extremal surfaces. Since extremal
surfaces have minimal area we are looking to minimize the following:

\begin{equation}
\lambda= \int 2 \pi \rho \Psi^2 [ \Psi^4 dz^2 + \Psi^4 d\rho^2] ^{1/2} \end{equation}

This can be rewritten as:

\begin{equation}
\lambda= \int 2 \pi \rho \Psi^2 [ \Psi^4 \left(\frac{dz}{d\sigma}\right)^2 + \Psi^4 \left( \frac{d\rho}{d\sigma}\right)^2] ^{1/2} d\sigma
\end{equation}

 After the following transformation $Q = \rho \Psi^4$ we obtain:

\begin{equation}
\lambda= \int 2 \pi [Q^2 \left(\frac{dz}{d\sigma}\right)^2 +  Q^2\left( \frac{d\rho}{d\sigma}\right)^2] ^{1/2} d\sigma\end{equation}

Letting $\frac{d}{d\sigma} = \dot{} \quad$ so that $L = (Q^2\dot{z}^2 +Q^2\dot{\rho}^2 )^{1/2}$, we can use Euler-Lagrange equation:

\begin{eqnarray}
L_z &=& \frac{d}{d\sigma}\left[ L_{\dot{z}} \right] \nonumber\\
\frac{1}{L}QQ,_{z} (\dot{z}^2 +\dot{\rho}^2)& =& \frac{d}{d\sigma} \left [\frac{1}{L}Q^2 \dot{z} \right]
\label{Euler}
\end{eqnarray}

Note that $\frac{1}{L} = \frac{d\sigma}{d\lambda}$. Multiplying equation \ref{Euler} by $\frac{d\sigma}{d\lambda}$ gives:

\begin{eqnarray}
 \frac{d\sigma}{d\lambda} \left[
\frac{1}{L}QQ,_{z} \left [ \left( \frac{dz}{d\sigma} \right) ^2+ \left( \frac{d\rho}{d\sigma} \right) ^2 \right] \right] & =&  \frac{d\sigma}{d\lambda}  \left [ \frac{d}{d\sigma} \left [\frac{1}{L}Q^2 \frac{dz}{d\sigma} \right] \right] \nonumber\\
\left(\frac{d\sigma}{d\lambda} \right)^2QQ,_z \left [ \left( \frac{dz}{d\sigma} \right) ^2+ \left( \frac{d\rho}{d\sigma} \right) ^2 \right] &=& \frac{d}{d\lambda} \left[ Q^2 \frac{dz}{d\lambda} \right] \nonumber \\
QQ,_z \left [ \left( \frac{dz}{d\lambda} \right) ^2+ \left( \frac{d\rho}{d\lambda} \right) ^2 \right] &=& \frac{d}{d\lambda} \left[ Q^2 \frac{dz}{d\lambda} \right]
\end{eqnarray}

Written in a different way:
\begin{eqnarray}
QQ,_z \left [ \dot{z} ^2+ \dot{\rho}^2 \right] &=&  \left[ Q^2 \dot{z} \right] \dot{} \nonumber \\
QQ,_z \left [ \dot{z} ^2+ \dot{\rho}^2 \right] &=&  2Q(Q,_z \dot{z} + Q,_{\rho} \dot{\rho}) \dot{z} + Q^2 \ddot{z}
\end{eqnarray}

Which gives the following equation:

\begin{equation}
Q \ddot{z} +Q,_z(\dot{z}^2+\dot{\rho}^2) +2Q,_{\rho} \dot{z}\dot{\rho}=0
\end{equation}

In a similar way the  second Euler-Lagrange equation:

\begin{equation}
L_{\rho} = \frac{d}{d\sigma}\left[ L_{\dot{\rho}} \right]
\label{G1}
\end{equation}

gives the following:

\begin{equation}
Q\ddot{\rho} + 2Q,_z\dot{z}\dot{\rho} + Q,_{\rho}(\dot{\rho}^2 - \dot{z}^2) = 0
\label{G2}
\end{equation}

Note also that the metric gives a first integral :

\begin{equation}
\dot{z}^2 + \dot{\rho}^2 = (\rho \Psi^4)^{-2}
\end{equation}

This allows the following parameterization for $z$ and $\rho$ in terms of $\lambda$.

\begin{equation}
\frac{dz}{d\lambda} = \frac{\cos{\alpha}}{\rho \Psi^4}, \quad    \frac{d\rho}{d\lambda} =  \frac{\sin{\alpha}}{\rho \Psi^4}
\end{equation}

Here $\alpha$ represents the direction of the trajectory of a ray
moving in the $(\rho,z)$ plane \cite{Bishop84}.  With this new
representation the geodesic equations (\ref{G1}, \ref{G2}) can be
summarized as a system of three ordinary differential equations. These
equations, when solved numerically, describe the path of light rays
moving in the hypersurface:

\begin{eqnarray}
 \frac{d\rho^2}{d\lambda} &=& \frac{2 \sin{\alpha}}{\Psi^4}\\
 \frac{d z\rho^2}{d\lambda}& =& \frac{\rho \cos{\alpha} +  2 z \sin{\alpha}}{\Psi^4}\\
 \frac{d \alpha \rho^2}{d\lambda} &= & \frac{\cos{\alpha}}{\Psi^4}(1+ 4 \rho \frac{\Psi,\rho}{\Psi}) + \frac{\sin{\alpha}}{\Psi^4}( 2 \alpha - 4 \rho \frac{\Psi,z}{\Psi})\nonumber\\
 \label{Bishopeq}
\end{eqnarray}

\begin{figure}[h]
\begin{center}
        \includegraphics[angle=0,width=4in, trim =130 310 180 110, clip=true]{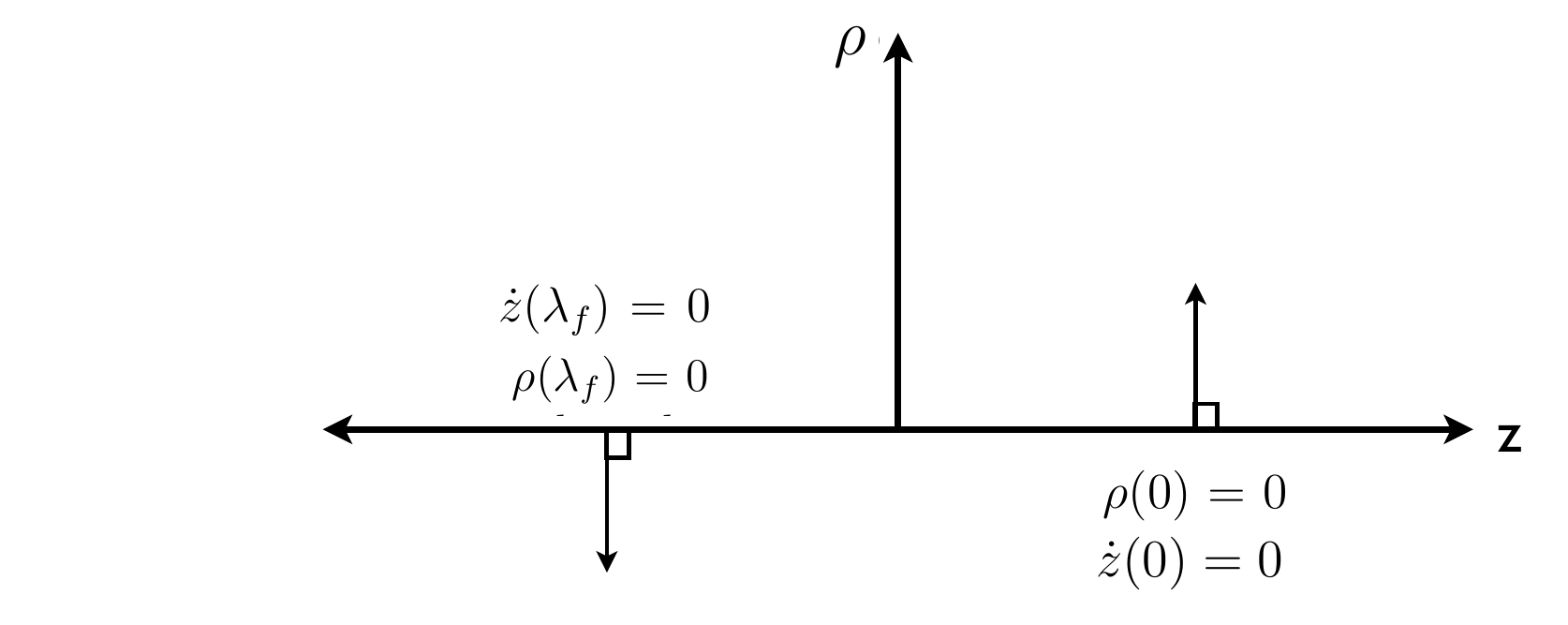}\\
        \caption{Boundary Conditions for a system of $N$ black holes}
 \end{center}
 \end{figure}

Marginally outer-trapped surfaces are represented by those rays that
start perpendicularly and end perpendicularly to the z axis. This
means: $\rho(0)=0 \rho(\lambda_f)=0$ and $\dot{z}(0)=
\dot{z}(\lambda_f)= 0$. Where $\lambda_f$ represents the value of the
parameter $\lambda$ when the ray returns to the z axis.
\newline

\subsection{Numerical Methods}

The system of three ordinary differential equations was solved using
{\it Mathematica} (for a description of the code see Appendix
B). To improve speed, the equations were rewritten using the following
transformations:

\begin{equation}
A= \rho^2,  \; B= z \rho^2,\;  C=\alpha \rho^2
\end{equation}

Which gives the following system of equations:\\

\begin{eqnarray}
\displaystyle \frac{ dA}{d\lambda} &= & \frac{2 \sin{(\frac{C}{A})}}{\Psi^4}\\
\displaystyle \frac{dB}{d\lambda} & =& \frac {\cos{(\frac{C}{A})\sqrt{A} }}{\Psi^4} + \frac{2 \frac{B}{A} \sin{(\frac{C}{A})}}{\Psi^4}\\
\displaystyle \frac{dC}{d\lambda} & = & \frac{\cos{(\frac{C}{A})}}{\Psi^4}(1 + 4  \frac{\Psi,\rho}{\Psi}\sqrt{A}) + 
\frac{\sin {(\frac{C}{A})}}{\Psi^4} ( 2\frac{C}{A} - 4  \frac{\Psi,z}{\Psi}\sqrt{A})
\label{NumericalODE}
\end{eqnarray}

\vspace{8 mm}

With initial conditions:\\

\begin{equation}
A(0) = 0, \; B(0) =  0, \; C(0)=0
\end{equation}

In order to avoid division by zero, due to the initial conditions
$z(0)=z_o$ and $\rho(0)=0$, a Taylor expansion was used to rewrite the
initial conditions for the new variables A,B,C .

 \begin{equation}
 A(0) =  2 \lambda_o , B(0) = 2 \lambda_o z_o, C(0)= \pi \lambda_o
 \end{equation}

With $\lambda_o = 10 ^{-12}$.\\

\subsection{Procedures}

When the total mass of the system is distributed so that each black
hole has the same mass, the MOTS are symmetric with respect to the
$\rho$ axis. This means that at $z=0$ the derivative of $\rho$ with
respect to $\lambda$ is zero ($\dot{\rho}=0$) and a numerical method,
such as the Bisection Method, can be used to determine the correct
initial condition $z_o$ that describes a MOTS (marginally outer
trapped surface). If $\dot{\rho}|_{z=0}\neq0$, then it can be
concluded that there are no MOTS for the given conditions.

\begin{figure}[h]
\begin{center}
        \includegraphics[angle=0,width=5in, trim = 300 210 100 280, clip=true]{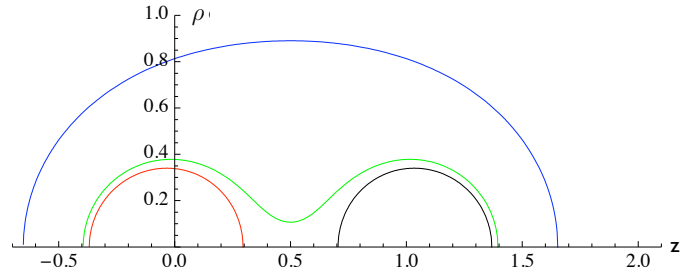}
        \caption{MOTS and Apparent Horizon for a system of two Black Holes}
        \label{MOTS}
         \end{center}
        \end{figure}

In the case of two black holes of different mass the above mentioned
method for finding the MOTS and apparent horizon does not apply. Since
the objective is to find the critical separation the method
implemented by Bishop \cite{Bishop84} can be used. Bishop found that
there are four different MOTS in a system of two black holes (figure
\ref{MOTS}). To find the critical distance the black holes are moved
farther apart until the two MOTS that surround both holes are joined
together. When this happens, the critical separation has been found.

For systems of three black holes distributed in a symmetrical manner
along the z axis, the critical separation is found by moving the
outermost black holes farther away from the origin until no outermost
MOTS is found.

\begin{figure}[h]
\begin{center}
        \includegraphics[angle=0,width=4in, trim = 50 270 120 160, clip=true]{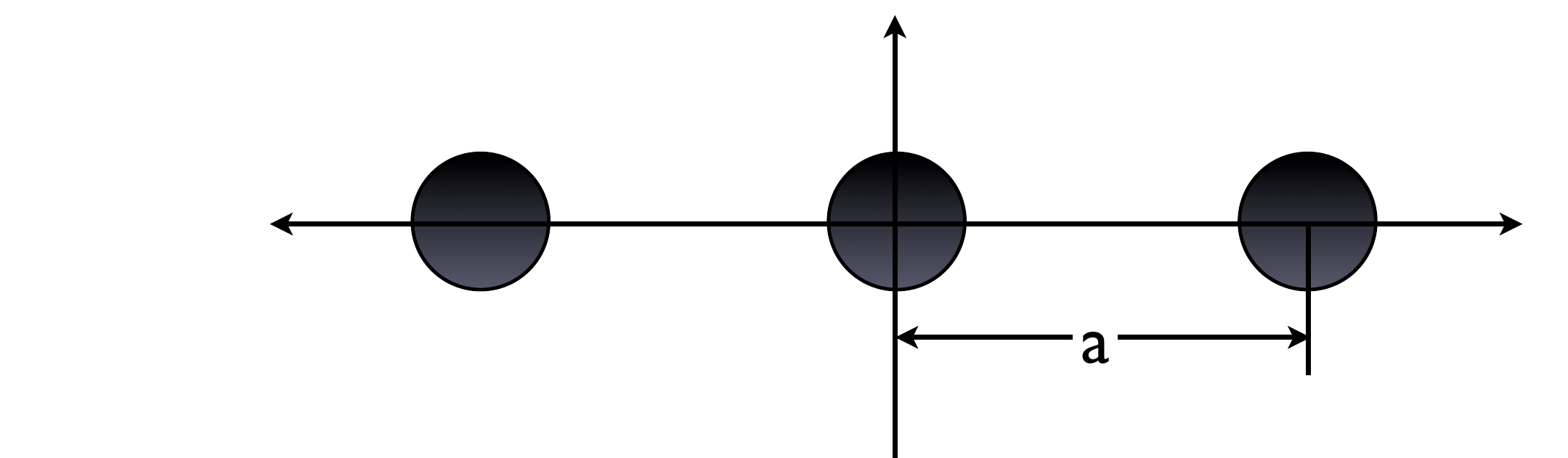}\\
        \caption{System of three Black Holes}
	\label{3BH}
\end{center}
\end{figure}

\begin{figure}[h]
\begin{center}
        \includegraphics[angle=0,width=4in, trim = 0 270 70 150, clip=true]{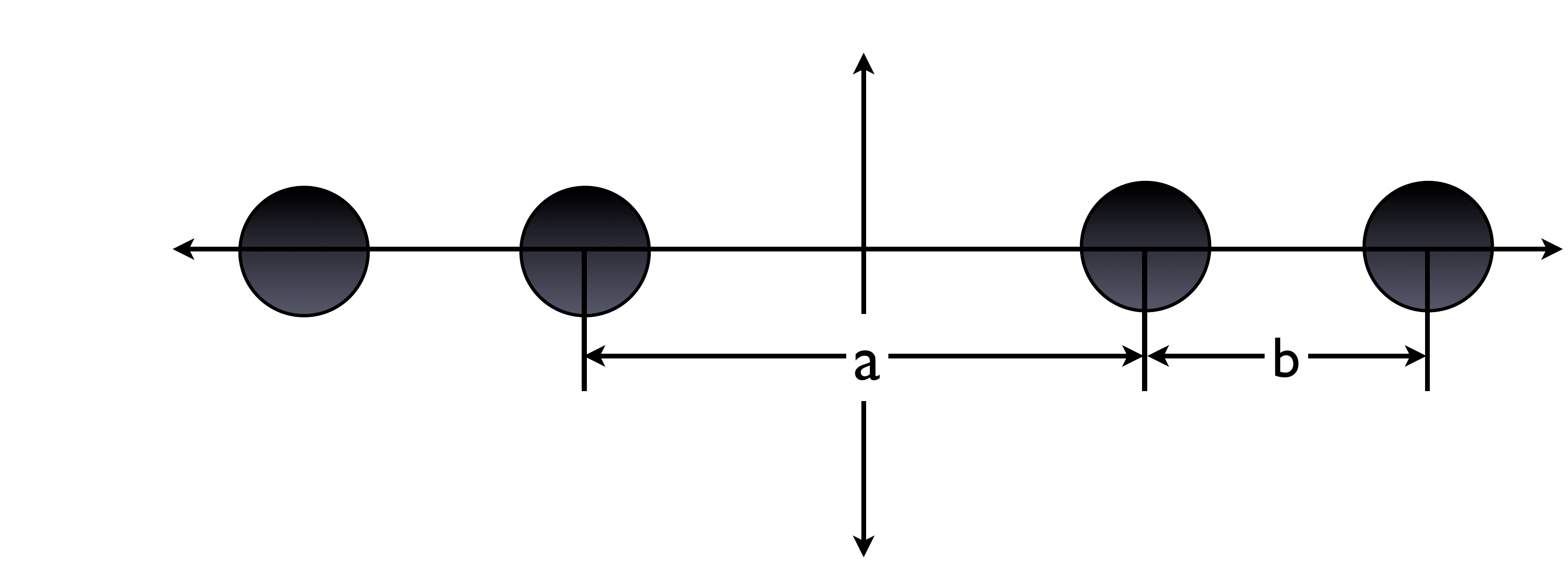}\\
        \caption{System of four Black Holes}
        \label{4BH}
 \end{center}
 \end{figure}

In the case were the system has four black holes there are two
distances that need to be taken into consideration. The distance
between the inner black holes, defined here as $a$, and the distance
between the outermost and inner black hole, defined here as $b$
(figure \ref{4BH}). In this case the critical values $a$ and $b$ are
found by first finding the position of the outermost black holes that
is farthest away from the origin ($f_{max} =a/2 + b $) and then moving
the inner black holes farther away until the largest value for $a$ is
found with its corresponding value for $b$.

\begin{figure}[h]
\begin{center}
        \includegraphics[angle=0,width=4in, trim = 0 270 70 150, clip=true]{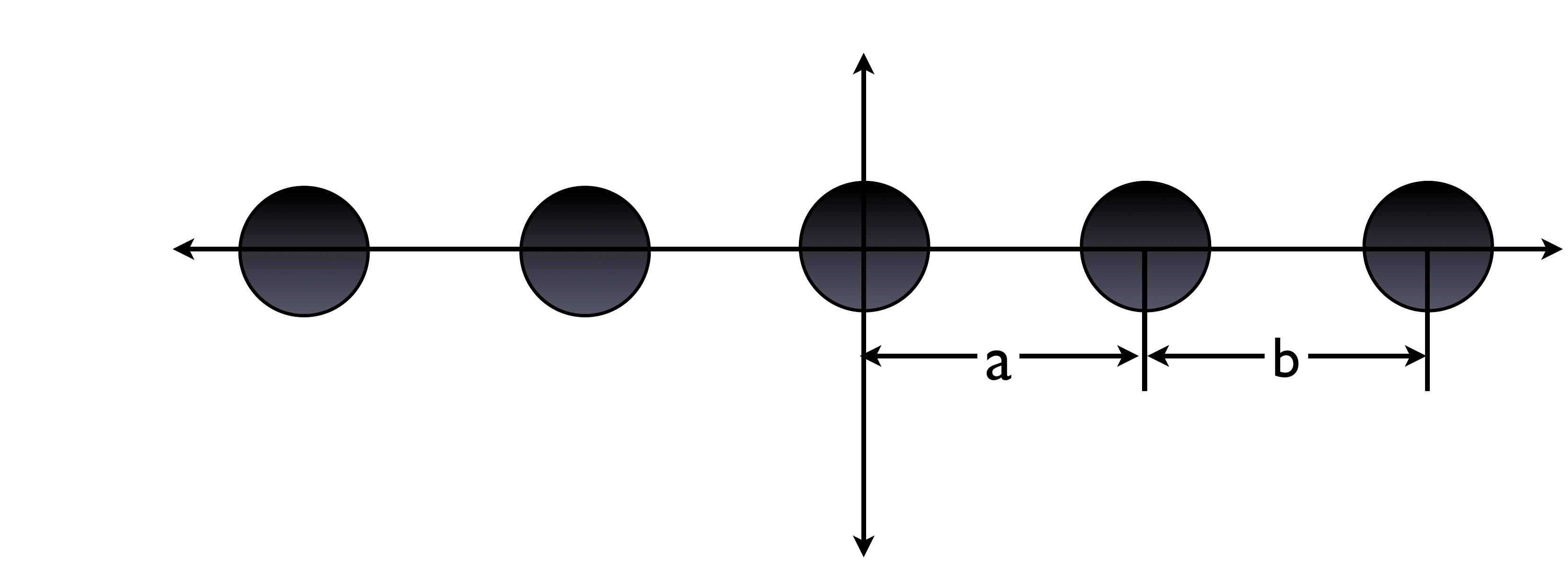}\\
        \caption{System of five Black Holes}
        \label{5BH}
        \end{center}
        \end{figure}

The same method is used for a system of five black holes. The variable
$a$ is now defined as the distance between the black hole located at
the origin and either of the adjacent black holes, which are here
referred to as inner black holes. The distance between the inner and
outermost black hole is defined as $b$ (figure \ref{5BH}). Finding the
critical separation is similar to the previous case of four black
holes, but now $f_{max} = a +b$.

\begin{figure}[h]
\begin{center}
\includegraphics[angle=0,width=3.5in, trim= 0 0 10 0, clip=true]{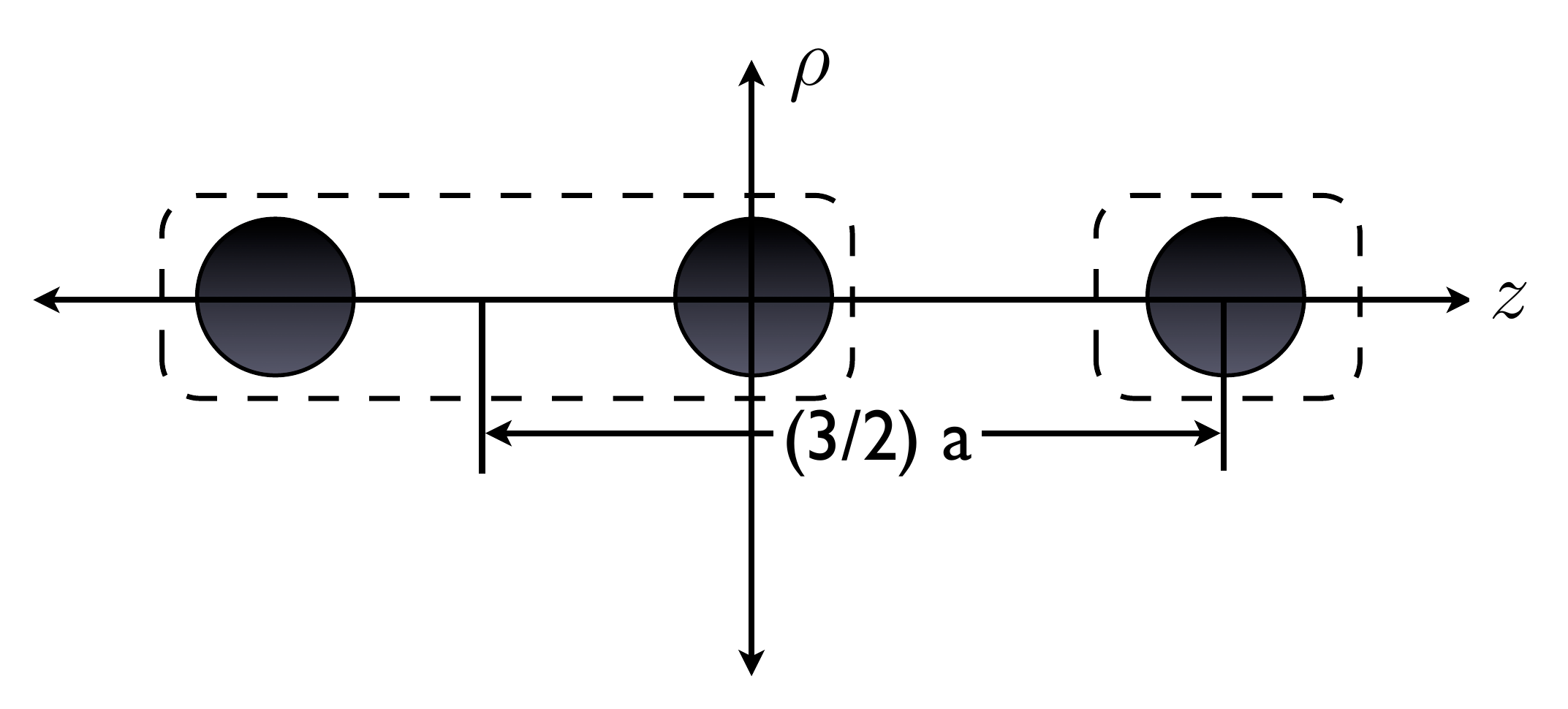}
\caption{Three black holes represented as two black holes with a mass ratio of 2:1}
\label{3BH2BH}
\end{center}
\end{figure}

For comparative reasons the black holes in each system are
hypothetically grouped together in order to model the system as a two
black hole system. This means that the black holes are assumed to be
grouped in such a way that they would form two clusters. For example,
in a system of three black holes we can put two black holes together
and leave the third one by itself. This grouping results in a system
of two black holes with a mass ratio of $2:1$ and a critical
separation $a_c = 1.5a$ (figure \ref{3BH2BH}).

\begin{figure}[h]
  \begin{center}
\includegraphics[angle=0,width=3.5in, trim= 0 0 10 0, clip=true]{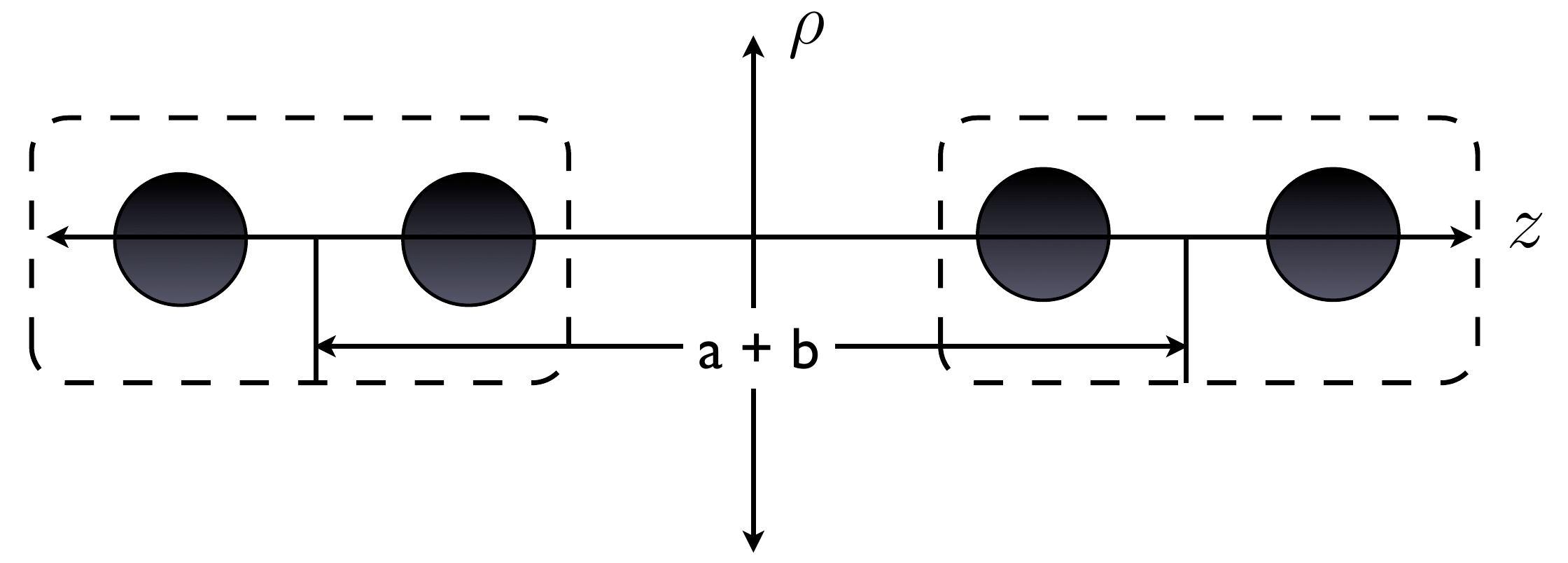}
\caption{Four black holes represented as two black holes with a mass ratio of 1:1}
\label{4BH2BH1}
\end{center}
\end{figure}

\begin{figure}[h]
 \begin{center}
\includegraphics[angle=0,width=3.5in, trim= 0 0 10 0, clip=true]{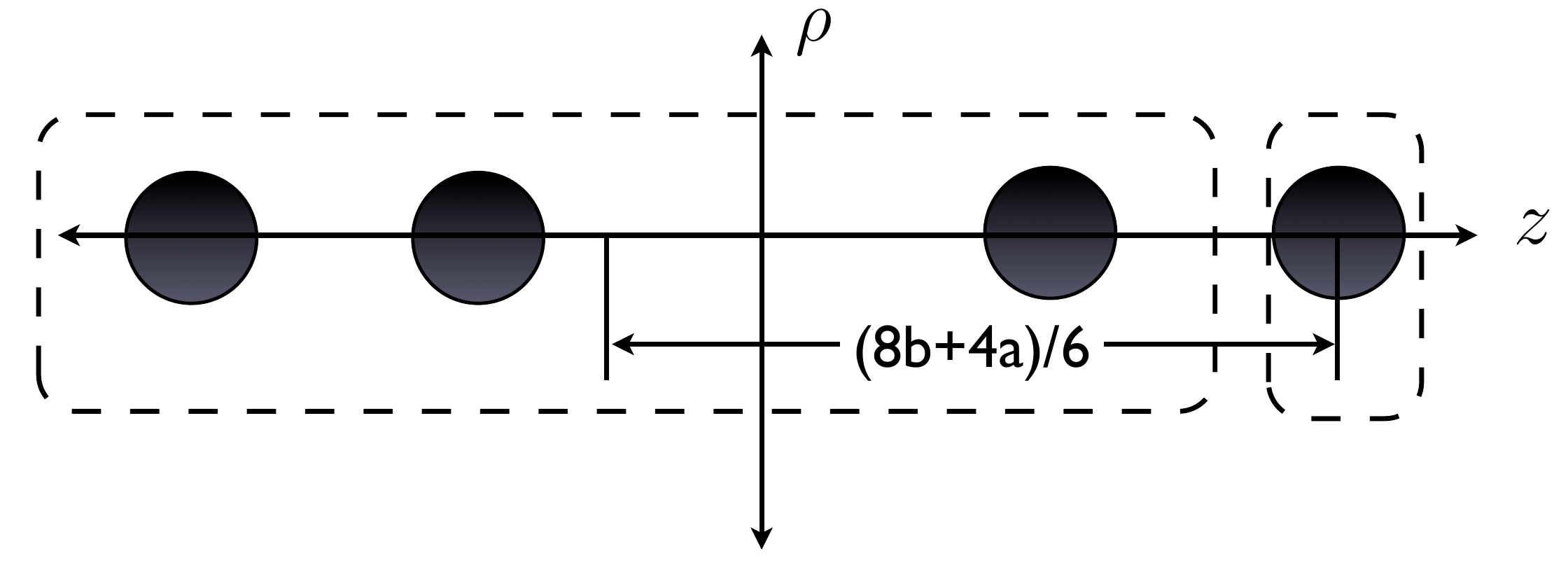}
\caption{Four black holes represented as two black holes with mass ratio of 3:1}
\label{4BH2BH2}
\end{center}
\end{figure}

The system of four black holes has two representations, one as a
system of two black holes with a mass ratio $1:1$ and a critical
separation equal to $a_c=a+b$ (figure \ref{4BH2BH1}), and a second one
as a system of two black holes with a mass ratio of $3:1$ and a
critical separation of $a_c = \frac{4a+8b}{6}$ (figure
\ref{4BH2BH2}).\\

\begin{figure}[h]
\begin{center}
\includegraphics[angle=0,width=3.5in, trim= 0 0 10 0, clip=true]{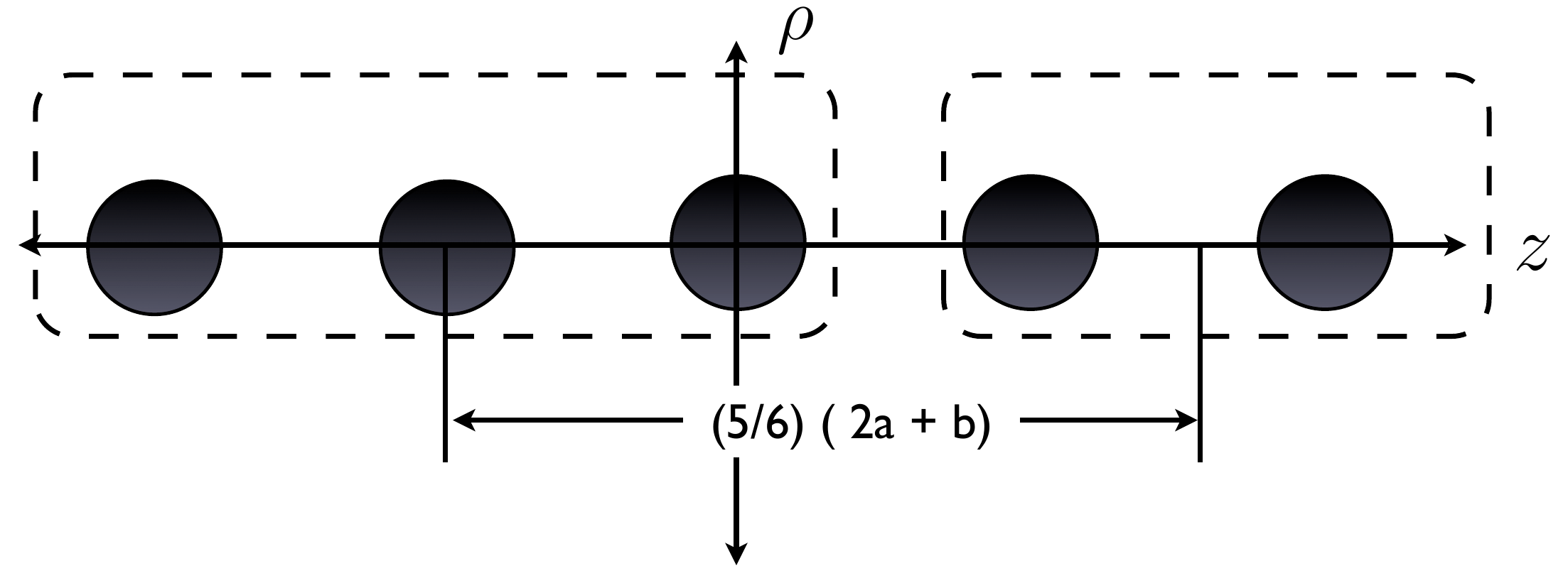}
\caption{Five black holes represented as two black holes with a mass ratio of 3:2}
\label{5BH2BH1}
\end{center}
\end{figure}

\begin{figure}[h]
\begin{center}
\includegraphics[angle=0,width=3.5in, trim= 0 0 10 0, clip=true]{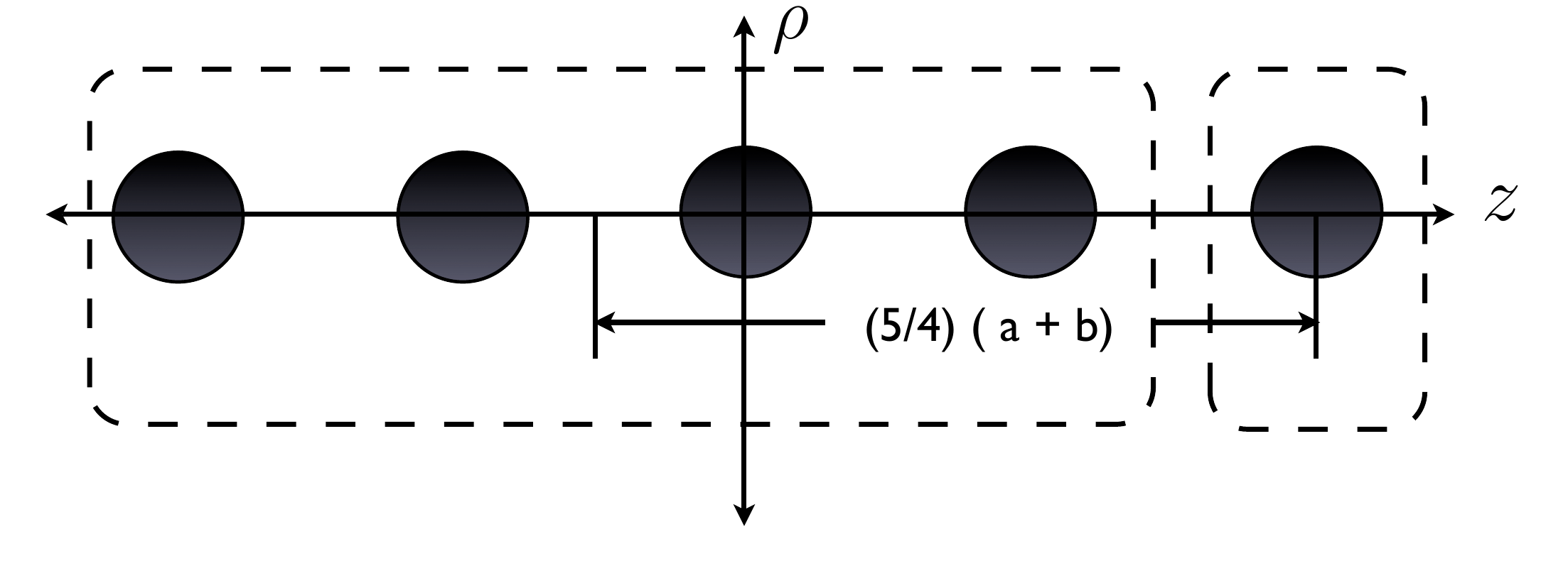}
\caption{Five black holes represented as two black holes with a mass ratio of 4:1}
\label{5BH2BH2}
\end{center}
\end{figure}

Finally the system of five black holes is represented as a system of
two black holes with a mass ratio $3:2$ and a critical separation
equal to $ a_c = \frac{5}{6} (2a +b)$ (figure \ref{5BH2BH1}) and as a
system with a mass ratio $4:1$ and a critical separation equal to $a_c
= \frac{5}{4}(a+b)$ (figure \ref{5BH2BH2}).

\subsection{Results}

\begin{figure}[h]
\begin{center}
        \includegraphics[angle=0,width=4in, trim = 70 580 280 40, clip=true]{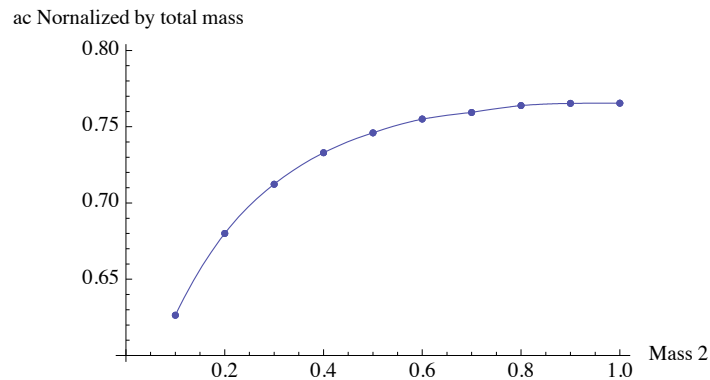}\\
        \caption{Plot of Critical separation normalized by mass vs. Mass of 2nd Black Hole}
        \label{CS2BH}
         \end{center}
           \end{figure}

In the case of two black holes with different mass the method
described in \cite{Bishop84} was implemented to relate the mass ratio
of the two black holes to the critical separation between them (table
\ref{2BHdmass} and figure \ref{CS2BH}). The table was used to predict
the critical separation for systems of $N$ black holes. To do this the
systems of $N$ black holes was first represented as systems of two
black holes. Then an equation for the critical separation was obtained
in terms of $a$ and $b$ (see figures \ref{3BH2BH} \ref{4BH2BH1}
\ref{4BH2BH2} \ref{5BH2BH1} and \ref{5BH2BH2}). Recall that depending
on the representation used, each system has a specific mass
ratio. Table \ref{2BHdmass} was used along with this ratio to find the
critical separation that corresponds to each case. This value was then
set equal to the equations for the critical separation and solved for
$a$ and $b$.

\begin{table}
\begin{center}
\begin{tabular}{| c | c | c |}
\hline
Mass $M_2$ & Critical Separation $a_c$& $a_c$ Normalized by total mass\\
\hline
1.0    & 1.531 & 0.7655\\
0.9 & 1.454 & 0.7653\\
0.8 & 1.375 & 0.7639\\
0.7 & 1.291 & 0.7594\\
0.6 & 1.208 & 0.7550\\
0.5 & 1.119 & 0.7460\\
0.4 & 1.026 & 0.7329\\
0.3 & 0.926 & 0.7123\\
0.2 & 0.816 & 0.6800\\
0.1 & 0.689 & 0.6264\\
\hline
\end{tabular}
\end{center}
\caption{ Two black holes of different mass ($M_1=1$ )}
\label{2BHdmass}
\end{table}

In the case of $N=3$ the mass ratio was $1:2$ and the critical
separation normalized by mass was $\frac{a_c}{3} = \left(\frac{3a}{2}
\right ) \left (\frac{1}{3} \right)= 0.746$ (see figure
\ref{3BH2BH}). Then:

\begin{eqnarray*}
\left(\frac{3a}{2} \right) \left( \frac{1}{3} \right) & = 0.7460\quad;\quad\quad a  =  1.492
\end{eqnarray*}

In the case of $N=4$ we can create two equations:

\[ \frac{a+b}{4} = 0.7655 \quad \textrm{ Mass ratio 1:1}\]
\[ \frac{8b +4a}{24} = 0.7203 \quad \textrm{Mass ratio 1:3}\]

Solving for $a$ and $b$ gives $ a = 1.8022$ and $b= 1.2598$. Finally in the case of $N=5$ we obtained:
\[ \frac{2a+b}{6} = 0.7582 \quad \textrm{Mass ratio 2:3}\]
\[ \frac{a+b}{4} = 0.6982 \quad \textrm{Mass ratio 1:4}\]
Solving for $a$ and $b$ gives $a = 1.7564$ and $b = 1.0364$.\\

The following table shows the results obtained for the critical
separations $a$ and $b$ for systems of two, three, four and five black
holes using the method described previously.

\begin{table}[h]
\begin{center}
\begin{tabular}{| c | c | c | }
\hline
$N^o$ Black Holes &$a$ &$b$\\
\hline
2& 1.531 & -- \\
3 & 1.528 &--\\
4 & 1.340 & 1.609 \\
5 & 1.370 & 1.650 \\
\hline
\end{tabular}
\end{center}
\caption{Critical Separations $a$ and $b$ for two, three, four and five black holes}
\label{CriticalS}
\end{table}

Comparing these results to the ones predicted by table \ref{2BHdmass}
gives the following errors:

\begin{table}[h]
\begin{center}
\begin{tabular}{| c | c | c | c | c | c | c |}
\hline
$N^o$ Black Holes & Numerical $a$ & Predicted $a$ & Error & Numerical $b$ & Predicted $b$ & Error\\
\hline
3 & 1.528 & 1.492 & 2.36\% & -- & -- & --\\
4 & 1.340 & 1.802 & 34.49\%&1.609 & 1.260& 21.70\% \\
5& 1.370 & 1.756 & 28.20\% & 1.650 & 1.036 & 37.19 \%\\
\hline
\end{tabular}
\end{center}
\caption{ Critical separations: Comparison between numerical results and predicted results}
\label{Error}
\end{table}

Were:
 \[ \textrm{Error}= \frac{100}{\textrm{Numerical}\quad a} (|\textrm{Numerical} \quad a - \textrm{Predicted} \quad a|)\]
 
\subsection{Discussion}

Note that these values are close to the ones predicted by table
\ref{2BHdmass} and can provide a good first guess for finding the
critical separations of systems of $N$ black holes. Although the
percentage error might seem large, when presented with the situation
of making a preliminary estimate for the values for these critical
separations in a system of $N$ black holes, which is useful
information when determining the location of the apparent horizon, any
estimate that is 20\% or 30\% of the actual value is reasonable.

This method can be extended to predict the location of the apparent
horizon for a system of any $N$ black holes symmetrically distributed by
following the these steps:

\begin{itemize}
\item[1.]Count the number of critical separations $a_i$. If $N$ is odd then the number of critical separations, $M$, is $\frac{N-1}{2}$ and if $N$ is even then the number of critical separations is $\frac{N}{2}$.
\item[2.] Establish all the possible distinct groupings of the $N$ black holes that would simulate a system of two black holes. The number of groupings should be equal to the number of critical separations $M$. 
\item[3.] For each grouping determine the location of the center of mass for the two clusters. Let $r_1$ ($r_2$) be the distance between the axis of symmetry and the center of mass of the left cluster (right cluster).
\item[4.] For each grouping find the mass ratio of the two clusters and using table \ref{2BHdmass} interpolate the critical separation $a_c$ that corresponds to that mass ratio.
\item[5.] Solve the system of equations given by $r_1+r_2 = a_c$ to find the values of all critical separations $a_1 \cdots a_M$.
\end{itemize}

By analyzing a system of two black holes we have been able to predict
the critical separations for system of multiple black holes. We have
developed a method that provides an adequate first approximation of
these critical separations and that if applied can significantly
reduce the time needed to find the apparent horizon by telling us if
we should be looking for a common apparent horizon that engulfs all
black holes, or if we should be looking for individual apparent
horizons surrounding each body.

\section{Black hole with a ring singularity}\label{ringBH}

The Motivation for studying black hole rings comes from computational
results from Shapiro et. al. \cite{Shapiro95a} in which the collapse of a
rotating toroidal configuration of collsionless particles to Kerr
black holes gives rise initially to an event horizon with toroidal
topology. The event horizon eventually becomes topologically
spherical. In this paper they explain that there is no violation of
topological censorship since when the toroidal horizon forms the
points in the inner rim of the torus (the whole of the torus) are
spacelike. This implies that the hole closes up faster than the speed
of light.

Their analysis begins with a two dimensional surface which has the
topology of an oblate spheroid. This surface will eventually represent
the event horizon after the black hole has reached its equilibrium
state. They trace back the light rays emanating in the normal
direction inward to the surface. The boundary of the spacetime points
in the casual past of this surface will be generated by the set of
light rays emanating from the surface that cross other light rays or
that focus to a point (that form a caustic).  They further explain
that in this case, where the initial surface is an oblate spheroid,
the rays that focus to a point will cross other light rays before they
form a caustic. So in essence the boundary of the casual past of this
surface is represented by the spacelike surface where all rays cross
(the crossover surface $X$). They have shown that this surface X has
toroidal topology.

They further explain that once the black hole has reached its
equilibrium state and the event horizon has its full complement of
generators then this horizon will have spherical topology (namely the
oblate spheroid represented by the above mentioned surface) agreeing
with theorems developed by Galloway and Browdy
\cite{Browdy:1995qu,Galloway:2006ws}.

What we want to do here is to use the apparent horizon as an approximation to
the event horizon. We will apply the previous method used for finding
the apparent horizon for systems of $N$ black holes to the case of a
black hole ring. This will allow us to find a specific mass of the
black hole ring that allows the formation of such toroidal event
horizon.

\subsection{Equations}

To adapt the equations developed by Bishop \cite{Bishop82} and used in
Sec. \ref{NBHs}, we first need to develop a new conformal factor that takes
into account the new circular shape of the black hole. To do so recall
that the conformal factor is given by:

\begin{equation}
 \Psi =  1 + \sum_i \frac{m_i}{2 R_i} 
\end{equation}

Where $R_i= r-r_i$ is the difference between a reference point $r
=(\rho,z)$ and the location of the $i^{th}$ black hole $r_i=(\rho_i,
z_i)$.\\

Consider a ring in the $z=0$ plane of radius $\rho=\rho_o$, then the
distance in cylindrical coordinates between any point in space ($\rho,
\varphi, z$) and the ring is given by S:

\begin{equation} 
S^2 = z^2 + ( \rho \cos{\varphi} -\rho_o \cos{\theta} )^2 + (\rho\sin{\varphi} - \rho_o \sin{\theta})^2
\end{equation}

Simplifying this expression we get:

\begin{equation}
S^2 = z^2 + \rho^2 +\rho_o^2 - 2 \rho \rho_o \cos{(\theta-\varphi)}
\end{equation}

Then the conformal factor for the metric is given by:

\begin{equation}
\Psi =1+\int_0^{2\pi} \frac{M}{2\,S}d\phi \;, \quad    \phi  = \theta  - \varphi
\label{confint}
\end{equation}

Here $M$ is the mass of the black hole ring.

Note that if the following conditions hold:
\begin{eqnarray}
Re[z^2 +(\rho-\rho_o)^2]>0&  \nonumber  \\
 Re[z^2+(\rho+\rho_o)^2]>0&  \nonumber \\
 \left| Re\left[ \frac{z^2+\rho^2+\rho_o^2}{\rho\rho_o}\right] \right| \geq 2  &\ \textrm{or}\ & \frac{z^2+\rho^2+\rho_o^2}{\rho\rho_o} \in \bbC 
 \end{eqnarray}

then:

\begin{equation}
\Psi =1+ \frac{M}{2} \left[ \frac{2 \textrm{EllipticK} [ \frac{-4 \rho\rho_o}{z^2+(\rho-\rho_o)^2}]}{\sqrt{z^2 + (\rho- \rho_o)^2}} + \frac{2\textrm{EllipticK}[\frac{4\rho\rho_o}{z^2+(\rho+\rho_o)^2}]}{\sqrt{z^2 +(\rho+\rho_o)^2}} \right] \\
\label{conformal}
\end{equation}

In {\it Mathematica} the EllipticK function
\footnote{In Maple the EllipticK function is defined in a different way,
essentially $\textrm{EllipticK}(x)\to \textrm{EllipticK}(\sqrt{x})$. Hence the Taylor expansion in this program is given by: 
\[ \frac{\pi}{2} + \frac{\pi x^2}{8} + \frac{9 \pi  x^4}{128}+\frac{25 \pi  x^6}{512}+\frac{1225 \pi  x^8}{32768}+\frac{3969 \pi  x^{10}}{131072}+\frac{53361 \pi  x^{12}}{2097152}+\frac{184041 \pi  x^{14}}{8388608}+O[x]^{16}\] 
This gives the following simplified form of the conformal factor:
\[\Psi = 1 +  \frac{M}{2}\left[\frac{4\,\textrm{EllipticK}\left( 2 \sqrt{\frac{\rho \rho_0}{z^2 +(\rho + \rho_o)^2}}\right)}{\sqrt{z^2 + ( \rho + \rho_o)^2}} \right] \]}
 is defined in such a way that its Taylor expansion around $x = 0$ gives:

\[\frac{\pi }{2}+\frac{\pi  x}{8}+\frac{9 \pi  x^2}{128}+\frac{25 \pi  x^3}{512}+\frac{1225 \pi  x^4}{32768}+\frac{3969 \pi  x^5}{131072}+\frac{53361 \pi  x^6}{2097152}+\frac{184041 \pi  x^7}{8388608}+O[x]^8\]

On the other hand if:
\begin{eqnarray}
 Im \left[ \frac{z^2+\rho^2+\rho_o^2}{\rho\rho_o} \right] =0  \quad& \textrm{and} &\quad \left |Re \left[ \frac{z^2+\rho^2+\rho_o^2}{\rho\rho_o} \right] \right| <2 \nonumber\\
 \textrm{or} \quad Re \left[ z^2 +(\rho-\rho_o)^2\right] &\leq&0  \nonumber \\
 \textrm{or} \quad  Re \left[ z^2 +(\rho +\rho_o)^2   \right] & \leq& 0 
 \label{condition2}
 \end{eqnarray}

then the integral in equation \ref{confint} can be performed. However, these last conditions will never hold since $z, \rho$ and $\rho_o$ are real numbers. A plot of the function:

\begin{equation}
f= \frac{z^2+\rho^2+\rho_o^2}{\rho\rho_o}
\end{equation}

rewritten using $ Z = \frac{z}{\rho}$ and $p = \frac{\rho_o}{\rho}$

\begin{equation}
f = \frac{Z^2 + p^2 +1}{p}
\label{cond2function}
\end{equation}

Shows that the expression $\left|Re \left[ \frac{z^2+\rho^2+\rho_o^2}{\rho\rho_o} \right] \right| <2$  will never hold:

\begin{figure}[h]
\begin{center}
        \includegraphics[angle=0,width=4in, trim =0 0 0 0, clip=true]{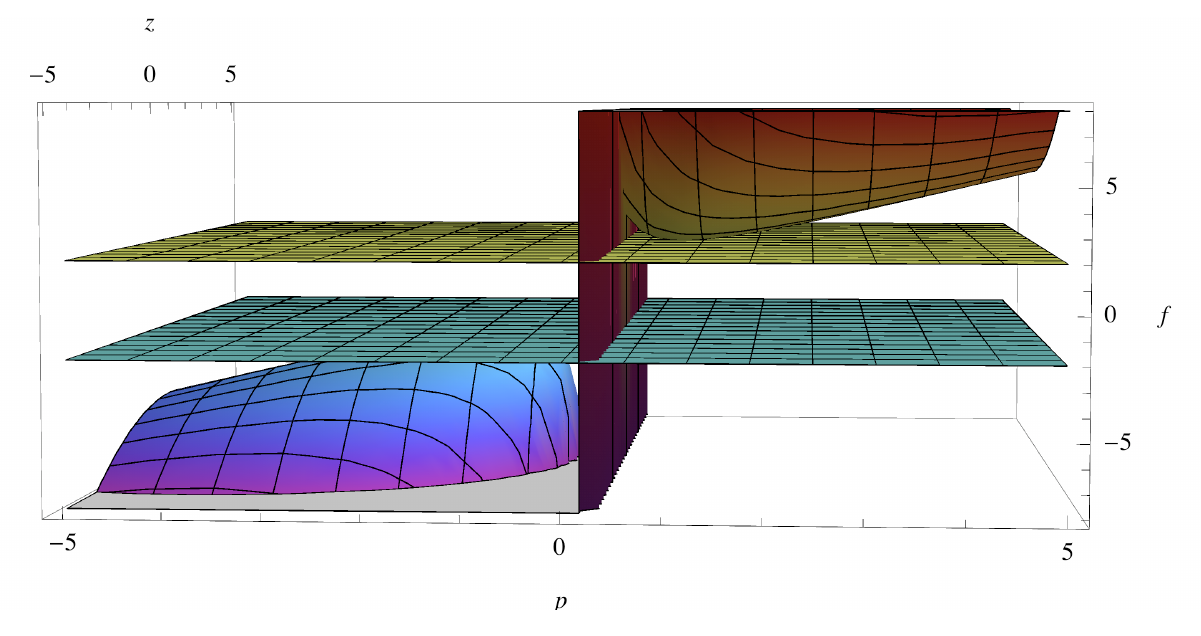}
        \caption{ Plot to show $\left|Re \left[ \frac{z^2+\rho^2+\rho_o^2}{\rho\rho_o} \right] \right| \not < 2$ }
        \end{center}
        \end{figure}
        
Hence the conformal factor $\Psi$ should be represented as in equation
\ref{conformal}. Used in conjunction with Bishop's equations
\ref{Bishopeq} we are able to find apparent horizons for black hole
rings.
 
\subsection{Procedures}

To find the apparent horizon, we again use equations
\ref{NumericalODE} and we start with the following initial conditions:
 
 \begin{equation}
A(0) = \rho_o, \quad B(0) = 0, \quad \alpha(0) = 0
\end{equation}

\begin{figure}[h]
\begin{center}
        \includegraphics[angle=0,width=3in, trim =0 0 0 0, clip=true]{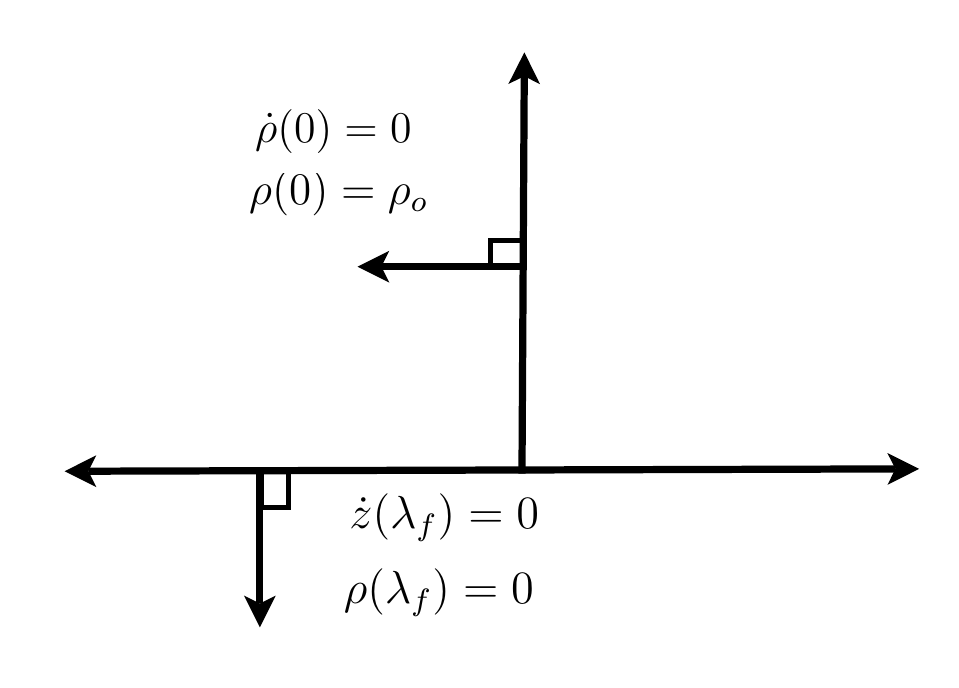}\\
        \caption{First set of boundary conditions for a black hole ring}
 \end{center}
 \label{FirstBC}
 \end{figure}
 
These initial conditions represent rays leaving perpendicular to the
$\rho$-axis ( $\dot{\rho}(0) = 0$) at the location $\rho(0)
=\rho_o$. We are interested in the rays that arrive perpendicular to
the $z$-axis since these rays will fulfill the boundary condition
$\dot{z}(\lambda_f) = 0$ (where $\lambda_f$ represents the value of
the parameter $\lambda$ when the ray returns to the $z$-axis) and
therefore they will represent the marginally outer trapped
surface. Unfortunately choosing to work in cylindrical coordinates to
account for the cylindrical symmetry does not allow these rays to
cross the $z$-axis and consequently we are not able to use of the
Bisection method to locate them accurately. We therefore choose to use
a visual method to find them. Since rays that are in the marginally
outer trapped surface never leave the surface, this means that these
rays will retrace their steps if the numerical integration code 
is left to run for a long
enough time. Hence we identify the apparent horizon with these rays.

Once this first approximation is obtained a new integration is
performed, this time using the same boundary conditions that we used
for finding the marginally trapped surfaces in the case of a system of
$N$ black holes:

\begin{equation}
A(0)=0, \quad B(0) =0, \quad C(0) =0
 \end{equation}
 Which need the same Taylor expansion as before, to avoid division by zero:
 
 \begin{equation}
 A(0) = 2 \lambda_o, \quad B(0)= 2 \lambda_oz_0, \quad C(0)= \pi \lambda_o 
 \end{equation}
 
 \begin{figure}[h]
\begin{center}
        \includegraphics[angle=0,width=4in, trim =130 310 180 110, clip=true]{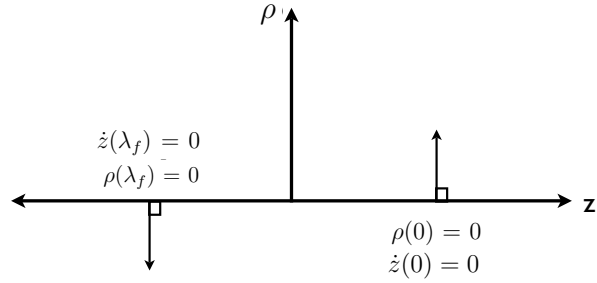}\\
        \caption{Second set of boundary conditions for a black hole ring}
 \end{center}
 \label{SecondBC}
 \end{figure}
 
The point $z(\lambda_f)$ in our first approximation, where the ray
reaches the $z$-axis perpendicularly, is going to be the starting
point for our second approximation. Now we can use the Bisection
method to find the apparent horizon. This means that we are looking
for rays that fulfill $\dot{\rho}=0$ at $z=0$.
   
\subsection{Results}
 
We present the some of the results obtained for the location of the
apparent horizon of a ring singularity of radius 1 in figure
\ref{Apparentone}, the rest are presented in appendix A. The graphs
show that as the mass decreases the apparent horizon becomes
compressed along the $z$-axis, consistent with the results observed in
the paper \cite{Shapiro95a}, where they find a
final event horizon with the topology of an oblate spheroid. This
results are better represented in table \ref{OblateResults}, which
shows the values obtained for the minor radius of the apparent
horizons $C_1$, their major radius $C_2$ and the ratio $C_1/C_2$.

\begin{table}
\begin{center}
\begin{tabular}{ |c | c| c|c|}
\hline
Mass $M$ & $C_1$ &$C_2$& $C_1/C_2$\\[0.5ex]
\hline
1.0$\pi$ & 6.226 & 6.311 & 0.987\\
0.8$\pi$& 4.955 & 5.062 & 0.979\\
0.6$\pi$& 3.674&3.816& 0.9623\\
0.5$\pi$ & 3.026 & 3.197& 0.946\\
0.4$\pi$& 2.367& 2.581&0.917\\
0.3$\pi$ & 1.683 & 1.972 & 0.853\\
0.25$\pi$ & 1.320 & 1.670 & 0.790\\
0.2$\pi$& 0.917 & 1.369 & 0.670\\
0.19$\pi$ & 0.825 & 1.307& 0.631\\
0.18$\pi$& 0.724 & 1.243 & 0.582\\
0.17$\pi$& 0.604 & 1.174 & 0.514\\
0.165$\pi$& 0.523&1.132 &0.462\\
0.163$\pi$& 0.478& 1.110& 0.430\\
\hline
\end{tabular}
\end{center}
\caption{ Results used for Extrapolation (Radius of black hole ring is 1)}
\label{OblateResults}
\end{table}

         \begin{figure}[h]
\begin{center}
        \includegraphics[angle=0,width=3in, trim =0 0 0 0, clip=true]{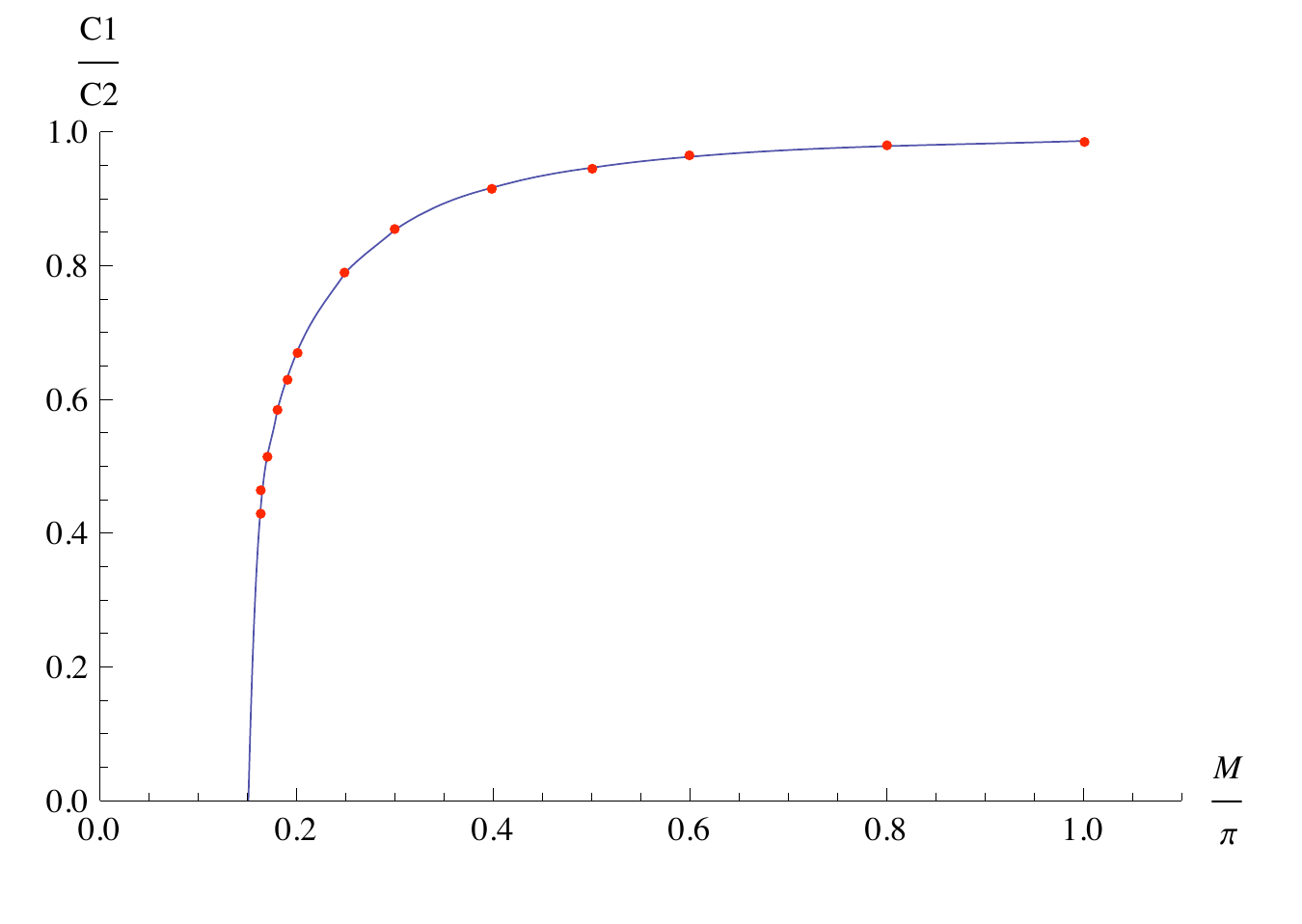}
        \caption{Plot of axis ratio $C_1/C_2$ as a function of mass $M/ \pi$}
        \label{C1C2}
        \end{center}
                \end{figure}

This table allowed us to establish a relation between the ratio
$C_1/C_2$ and the mass of the black hole ring. A plot of $C_1/C_2$
versus mass $M/ \pi$ is shown in figure \ref{C1C2}. Note how sharply
the ratio decreases once the mass of the black hole ring is less than
$M = 0.2 \pi $. Using the interpolation function from {\it
  Mathematica} we found that the mass that returns a ratio $C_1/C_2 =
0$ is $M=0.15 \pi$. Since there is an inverse relation between the
mass and radius of the ring, we can thus predict a critical radius
that will produce a toroidal event horizon using the value we obtained
for the mass. That is the critical radius is $R=1/(0.15 \pi) = 20/(3
\pi)$.

        \begin{figure}[h]
\begin{center}
        \includegraphics[angle=0,width=6in, trim =0 0 0 0, clip=true]{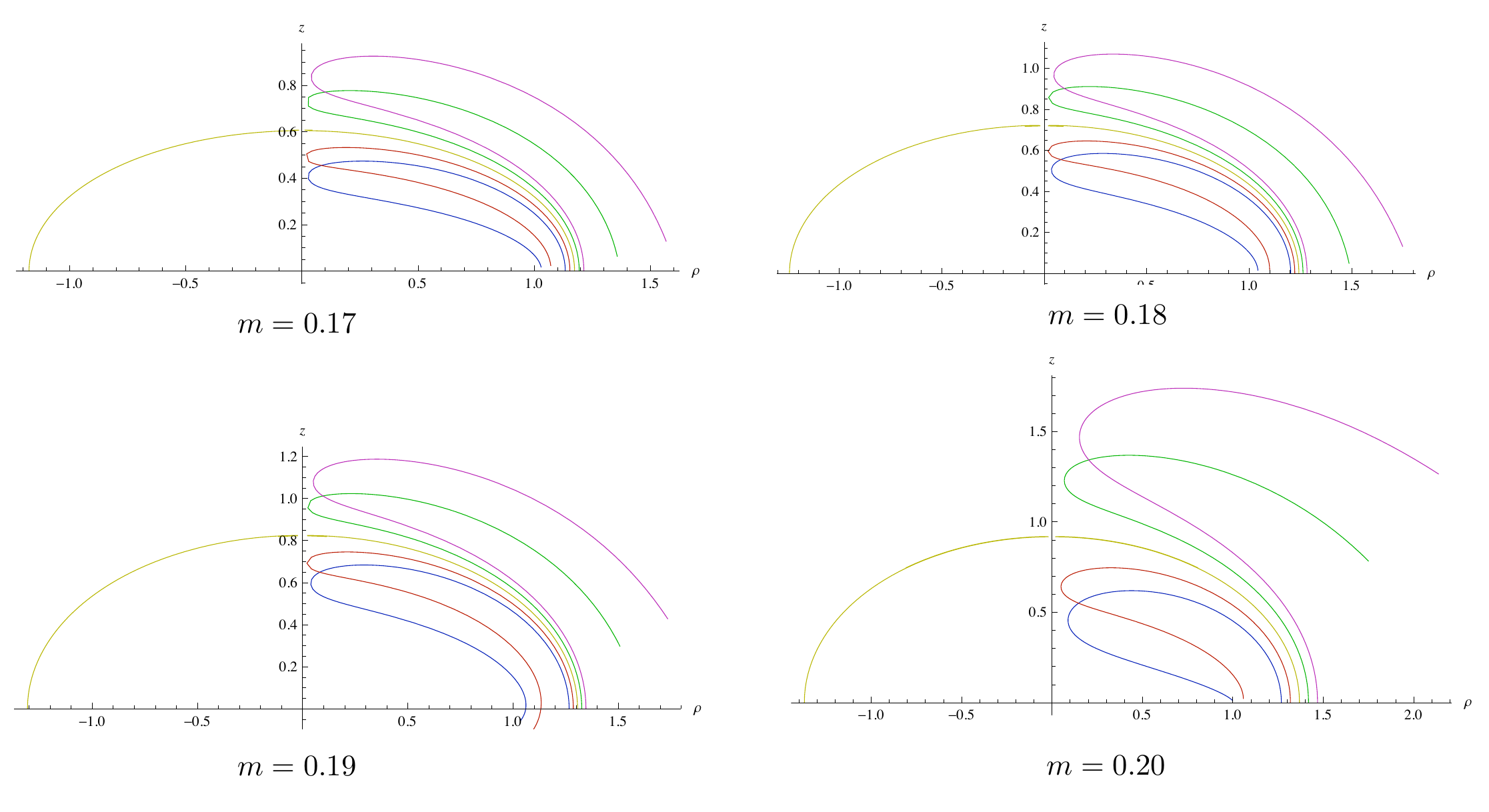}
        \caption{Apparent Horizon for ring masses $ m = 0.17, 0.18, 0.19, 0.20$ of radius 1}
        \end{center}
        \label{Apparentone}
        \end{figure}

\subsection{Discussion}

The main goal of this section was to develop a method for predicting
the size of a black hole ring that would give rise to an event horizon
of toroidal topology. This was accomplished by deducing the conformal
factor for a black hole ring and adapting the apparent horizon
equations found in
\cite{Bishop82} \cite{Bishop84} accordingly. The key argument here is
that even though an apparent horizon can never have toroidal topology
we can still use it to approximate the event horizon of black hole
rings that have spherical topology. The apparent horizon will follow
the shape of the event horizon up until it eventually becomes toroidal. So the
information we gathered for the flattening of the apparent horizon
can then be used to extrapolate the
value of the ring's mass that would give rise to a toroidal event
horizon. The results suggest that when the ring singularity has a mass
of $M=1 $ and a radius $R =20/(3 \pi) \approx 2.12$ (or equivalently
when the ring has a mass of $M = 0.15 \pi $ and a radius of $R=1$) the
event horizon would have toroidal topology.

\section{Conclusion}


As mentioned in the introduction apparent horizons are important in
numerical relativity because they provide a quasilocal boundary for
the black hole region. For instance, they are used in numerical
simulations to locate the black holes so that black hole excision
techniques can be used. They also provide physical information about
the black holes such as mass and angular momentum. With this in mind
and considering that recent full numerical research has focused on
systems of three black holes, 
we have focused our attention on gaining a better
understanding of systems of $N$ black holes.

To begin this analysis we focused on a time-symmetric spacelike
hypersurface with the purpose of developing a method for finding the
critical separations between the black holes in the system. This was
done by first analyzing a system of two black holes with different
mass and finding the critical separation for each mass ratio. The
result was a table that was used to predict the critical separations
for systems of $N$ black holes, represented as a system of two black
holes. This proved to be a good method for finding a first guess of
these critical separations. The errors obtained when comparing the
actual critical separation to the one predicted by the table were
around 20\% to 40\%. Although at first glance this errors seem large,
when confronted with a system of $N$ black holes, knowing whether to
look for a common apparent horizon or individual apparent horizons
makes a big difference on computational time.

Our next step was to consider a black hole ring. This was motivated by
papers which suggested the existence of event horizons of toroidal
topology in rotating clusters with toroidal topology. The equations
used to find the apparent horizon for the system of $N$ black holes were
adapted using a conformal factor that takes into account the circular
shape of the ring singularity. We vary its mass, while keeping its
radius constant, and computed its apparent horizon. The results were
apparent horizons with the topology of an oblate spheroid. A certain
minimal
mass was attained that did not allowed the formation of any spherical
apparent horizon suggesting that there is either no horizon or
the actual shape might be toroidal and
therefore not predictable by the algorithm. Using the data obtained we
constructed a table that relates the mass of the black hole ring to
the ratio of the minor radius to major radius of the apparent
horizon. Using this information we extrapolated the mass that
corresponds to a radius ratio equal to zero, thus suggesting that this
critical mass will correspond to a black hole ring with a toroidal
event horizon. Since there is an inverse relation between the mass and
radius of the ring we can alternatively, for a fixed mass of 1, find
the critical radius of the ring which in this case is $20/(3 \pi)
\approx 2.12$ M. While due to the smoothness of the {\it apparent} horizon
surfaces we cannot see a toroidal surface, it is interesting to
study the {\it event} horizon evolution for this configuration \cite{PLZ}.

A different way of constructing a toroidal horizon would be to consider
a set of $N$ black holes distributed along a circle at a critical separation
that connects all nearby horizons together. If one succeeds to do this on a
circle of radius 2.12 at least, with a total mass of 1, according to the
previous discussion we could create a toroidal horizon. In order to evaluate
this possibility with the apparent horizon information we have obtained
in Table \ref{CriticalS} 
we can study the progression of the critical length per mass
covered by a line distribution of $N$ black holes, representing an approximation
to an small portion of a ring.

Two black holes separated at a critical length $a_c$ will cover a 
length 
\begin{eqnarray}
\ell_c/M=(a_c/2+a_c+a_c/2)/N=2\,a_c/2=1.531 .\nonumber
\end{eqnarray}

For three black holes, see Fig.\ \ref{3BH}
\begin{eqnarray}
\ell_c/M=(a_c/2+2\,a_c+a_c/2)/N=3\,a_c/3=1.528 .\nonumber
\end{eqnarray}

For four black holes, see Fig.\ \ref{4BH}
\begin{eqnarray}
\ell_c/M=(b_c/2+b_c+a_c+b_c+b_c/2)/N=(a_c+3\,b_c)/4=1.54175 .\nonumber
\end{eqnarray}

And for five black holes, see Fig.\ \ref{5BH}
\begin{eqnarray}
\ell_c/M=(b_c/2+b_c+2\,a_c+b_c+b_c/2)/N=(2\,a_c+3\,b_c)/5=1.538 .\nonumber
\end{eqnarray}

So, essentially we cover 1.54 of a circle perimeter of unit mass, but we would
need to cover a perimeter of $2.12\times2\pi=13.32$. This leaves us
with a deficit factor of 8.65 to realize the toroidal horizon with
this construction.
\footnote{Note that the use of critical
distances in the conformal space as a 'physical' reference
are justified by the use of the our specific form of the initial data,
that in addition does not involve any choice of the slice in the form
of lapse and shift.}
However, since event horizons can show some fine
structure at the moment of merging, it is worth studying this configuration
in a more dynamical setting \cite{PLZ}.

In our search of common apparent horizons for rings of increasing
radius in Sec. \ref{ringBH}
we have not been able to find any beyond $R=2.12$. See Fig.\ \ref{C1C2}.
This leads to the question of the nature of the object left exposed without
a dressing horizon.
We recall the form of the conformal factor of the 3-metric
\begin{equation}
\Psi = 1 +  2M\left[\frac{\textrm{EllipticK}\left(\sqrt{\frac{2 \rho \rho_0}
{z^2 +(\rho + \rho_o)^2}}\right)}{\sqrt{z^2 + ( \rho + \rho_o)^2}} \right]
\end{equation}
Where this EllipticK function near the ring behaves like
\cite[page 591]{AbramowitzStegun},
\begin{equation}
\textrm{EllipticK}(\sqrt{x})\to\frac12\ln\left(\frac{16}{1-x}\right),
\end{equation}
for $x\to1$. This limit corresponds to approaching the ring as $\rho\to\rho_0$
and $z\to0$.
Upon double differentiation of the metric to compute the curvature
components, we find terms that diverge like $\ln^3|1-x|/|1-x|^2$. One can show that 
those effectively are true singularities of the spacetime computing,
for instance, scalar invariants \cite{PLZ}.

\bigskip\noindent
{\it Acknowledgments:}\\ 
The authors acknowledge important discussions
with M.Campanelli, B.Krishnan, M.Ponce, and Y.Zlochower.  We
gratefully acknowledge the NSF for financial support from Grants
No. PHY-0722315, No. PHY-0653303, No. PHY-0714388, No. PHY-0722703,
No. DMS-0820923, and No. PHY-0929114; and NASA for financial support
from NASA Grants No. 07-ATFP07-0158 and No. HST-AR-11763.
Computational resources were provided by the Ranger cluster at TACC
(Teragrid allocation TG-PHY060027N) and by NewHorizons at RIT.

\appendix

\section*{Appendix A: Apparent horizons for black hole rings} 

The following are the results obtained when finding the apparent horizon for a ring singularity.

\begin{figure}[h]
\begin{center}
        \includegraphics[angle=0,width=6in, trim =0 0 0 0, clip=true]{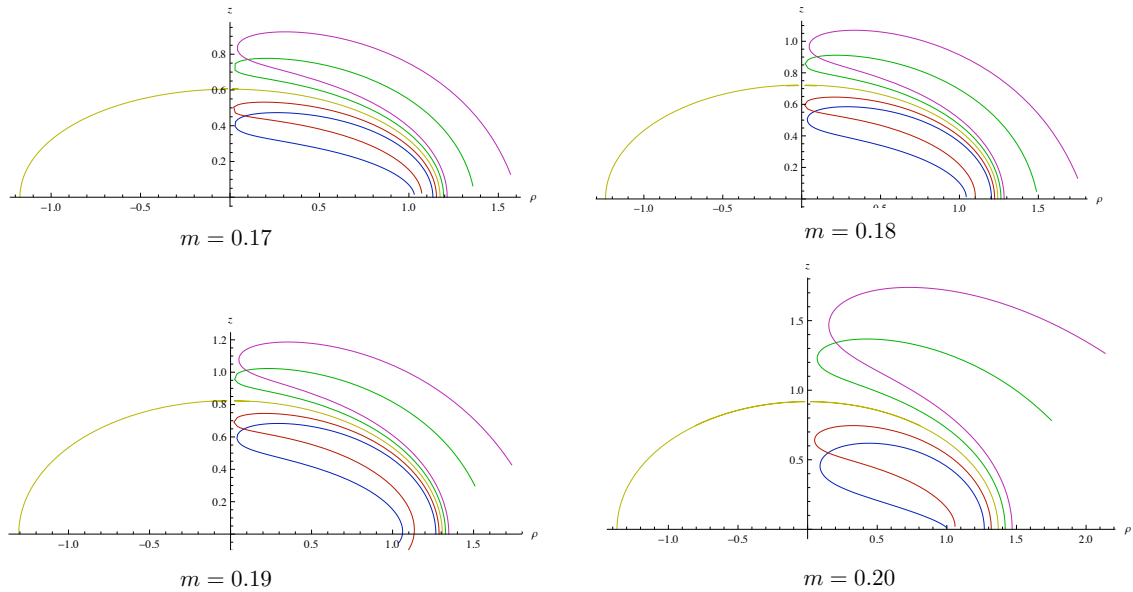}
        \caption{Apparent Horizon for a ring of radius 1 and masses $ m = 0.17, 0.18, 0.19, 0.20$}
        \end{center}

        \end{figure}
      
        \begin{figure}[h]
\begin{center}
        \includegraphics[angle=0,width=6in, trim =0 0 0 0, clip=true]{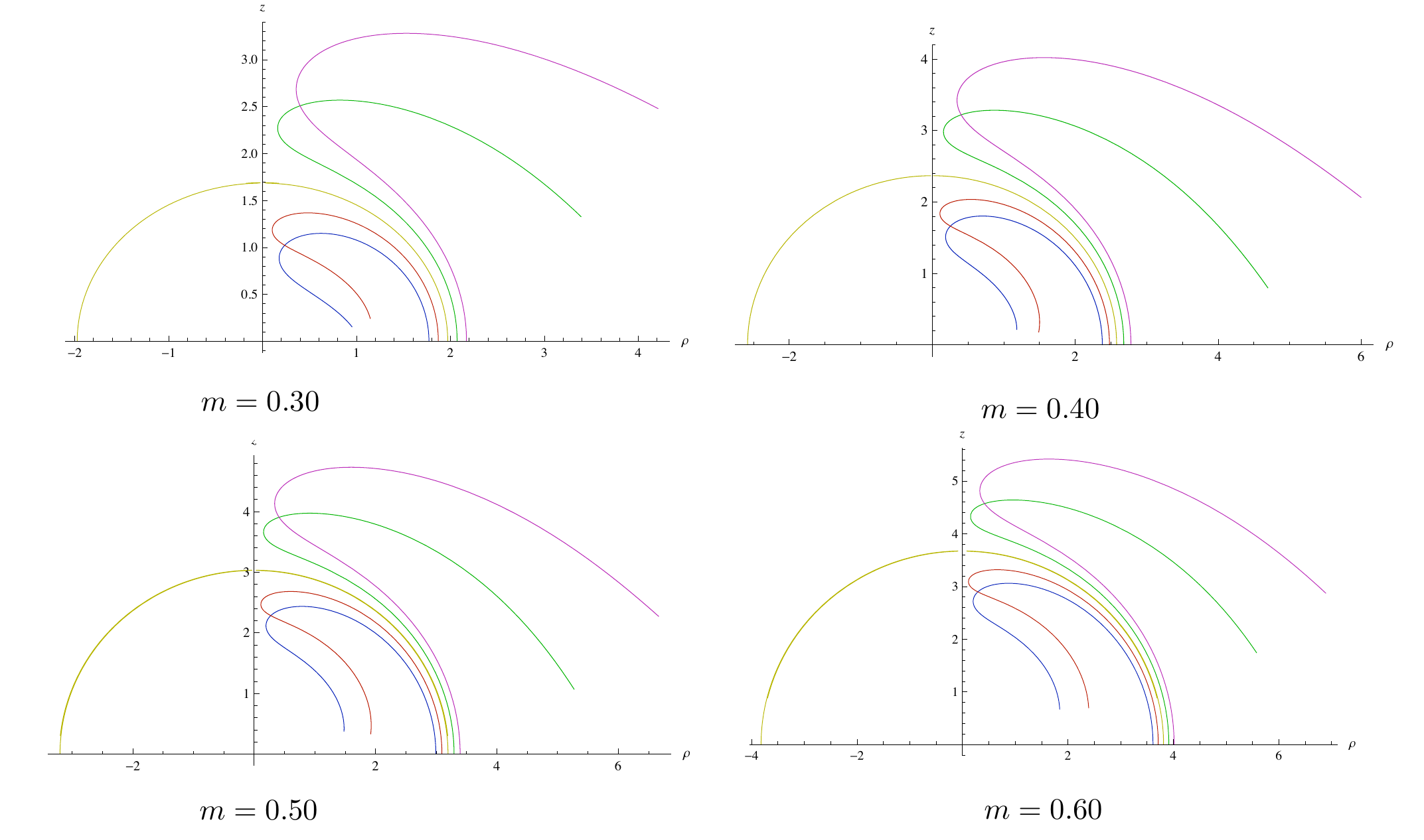}
        \caption{Apparent Horizon for a ring of radius 1 and masses $ m = 0.30, 0.40, 0.50, 0.60$}
        \end{center}
        
         \end{figure}  
        
         \vspace{3cm}
        \begin{figure}[h]
\begin{center}
        \includegraphics[angle=0,width=6in, trim =0 0 0 0, clip=true]{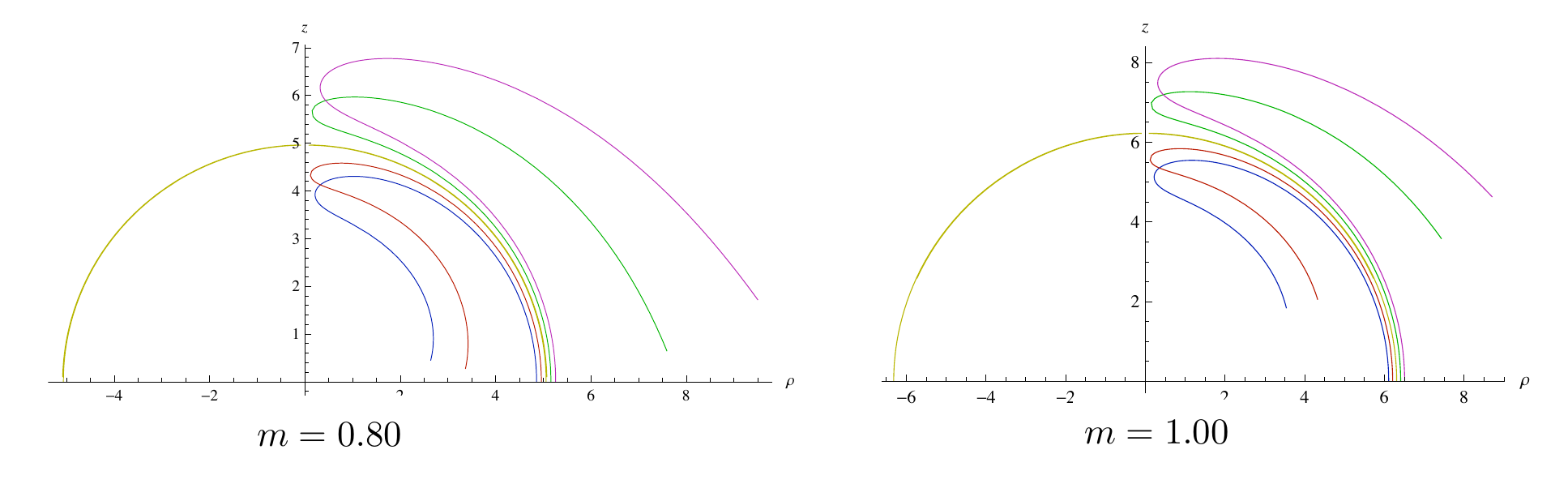}
        \caption{Apparent Horizon for a ring of radius 1 and masses $ m = 0.80, 1.00$}
        \end{center}
        
        \end{figure}

\section*{Appendix B: Data for systems of four and five black holes}

The first set of data was obtained when finding the apparent horizon
for four symmetrically distributed black holes. The distance $a$
represents the distance between the two inner black holes. The
distance $b$ represents the distance between the outer black holes and
the inner black holes. The second set of data was obtained when
finding the apparent horizon for five symmetrically distributed black
holes. The value $a$ represents the distance between the middle black
hole and the inner black holes. The distance $b$ represents the
distance between the outer black holes and the inner black holes.

\begin{center}
\begin{figure}[h]
        \includegraphics[angle=0,width=6in, trim =0 0 0 0, clip=true]{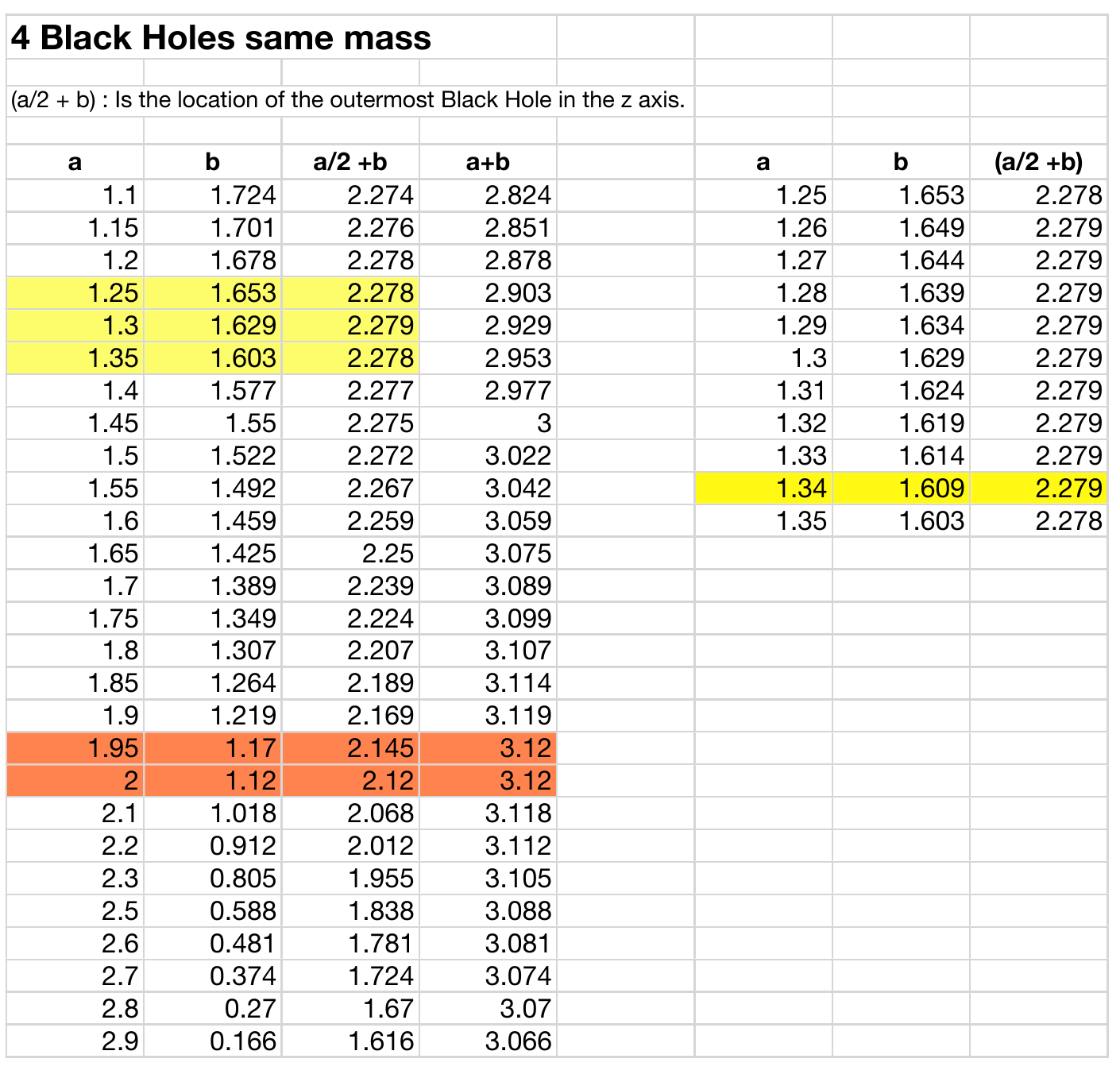}\\
                \caption{Data obtained for a system of 4 black holes}
                                \end{figure}
            \end{center}
            
            \begin{center}
\begin{figure}[h]
        \includegraphics[angle=0,width=6.5in, trim =0 0 0 0, clip=true]{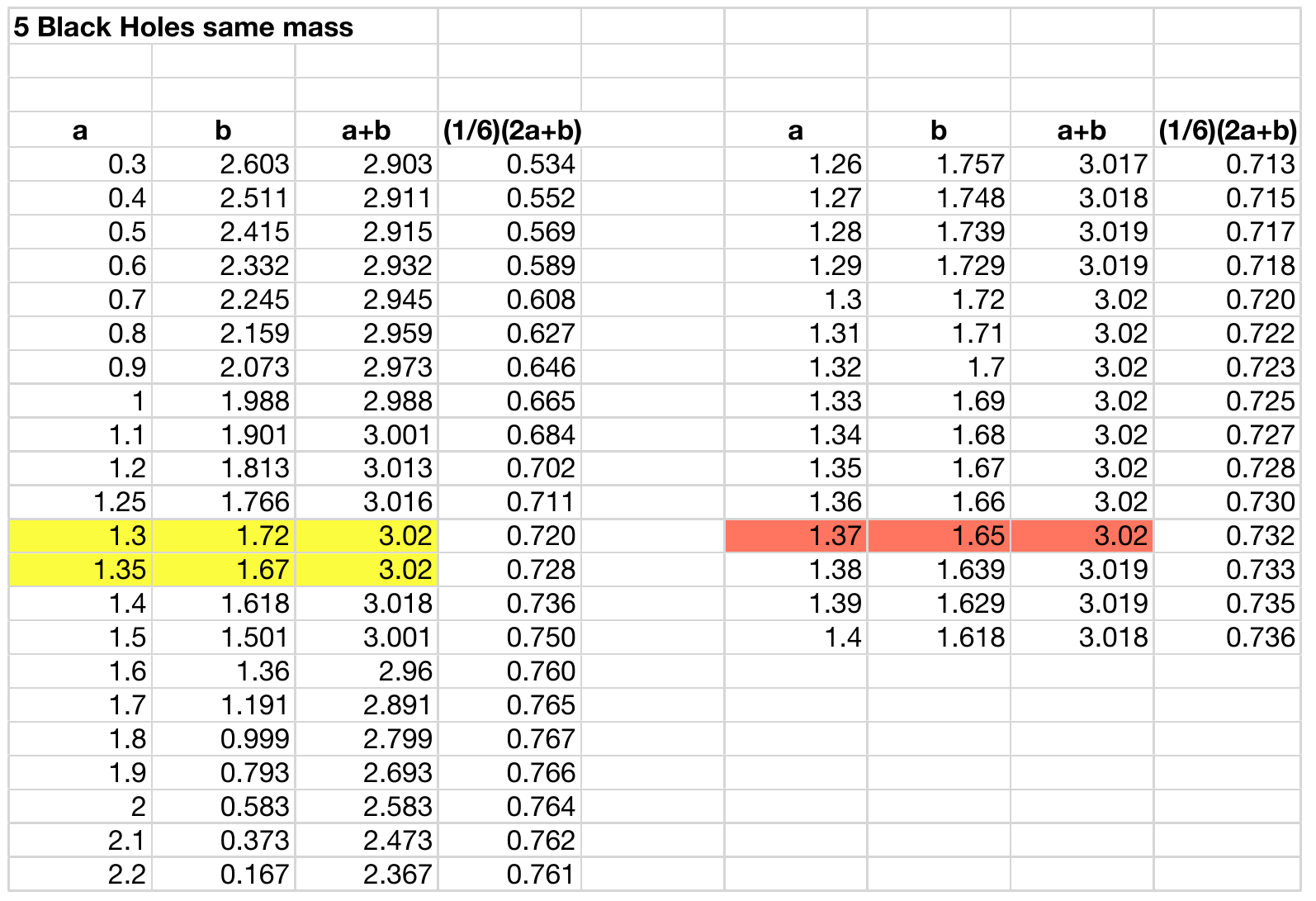}\\
               \caption{Data obtained for a system of 5 black holes}
                \end{figure}
            \end{center}

\section*{Appendix C: Code used for finding apparent horizons}

For a detailed description of the {\bf NDSolve} command from {\it Mathematica}, which was used to solve the system of non linear ODE's,  please refer to :\\

{ \it http://reference.wolfram.com/mathematica/ref/NDSolve.html }\\

The method used for the integration was an extrapolation method. This
method was chosen because, as explained in the {\it Mathematica}
documentation, it is an arbitrary order method that has automatic
order and step size controls. The arbitrary order means that they can
be arbitrarily faster than fixed-order methods for very precise
tolerances. A more detailed description of extrapolation methods can
be found in \cite{Stoer-Bulirsch-1980}. 
The sub-method used is a linearly implicit
Euler method (Also known as backward Euler method). For more
information the following website contains a complete description of
the extrapolation method.\\

{\it http://reference.wolfram.com/mathematica/tutorial/NDSolveExtrapolation.html}

\section*{References}

\bibliographystyle{unsrt}
\bibliography{../../../Bibtex/references}

\end{document}